\documentclass[10pt]{article}
\usepackage{amsmath,amssymb,mathtools,amsthm,latexsym}
\newcommand{\half}{\frac{1}{2}}
\newcommand{\ud}{\mathrm{d}}
\newcommand{\der}[3][]{\frac{\ud^{#1} #2}{\ud #3^{#1}}}

\newcommand{\Sl}{\mathrm{sl}}
\newcommand{\R}{\mathbb{R}}

\author{G.~Gubbiotti$^1$ \and M.~C.~Nucci$^2$}
\title{Superintegrable systems in non-Euclidean plane: hidden symmetries leading to linearity}
\date{$^1$School of Mathematics and Statistics F07, University of Sydney NSW 2006, Australia\\ [0.2cm]
$^2$Dipartimento di Matematica e Informatica, Universit\`a degli Studi di
Perugia, \& INFN Sezione di Perugia, 06123 Perugia, Italy}

\begin{document}

\maketitle

\begin{abstract}
Nineteen classical superintegrable systems in two-dimensional non-Euclidean spaces
    are shown to possess hidden symmetries leading to their linearization. They are the two Perlick systems [A. Ballesteros, A. Enciso, F.J. Herranz and O. Ragnisco,
{\em Class. Quantum Grav. 25}, 165005 (2008)], the Taub-NUT system [A. Ballesteros, A. Enciso, F.J. Herranz, O. Ragnisco,  and D. Riglioni, {\em SIGMA 7}, 048 (2011)],
 and all the seventeen superintegrable systems for
 the four types of Darboux spaces as determined in [E.G. Kalnins, J.M. Kress, W. Miller, P. Winternitz, {\em J. Math. Phys. 44}, 5811--5848 (2003)].
\end{abstract}

\section{Introduction}
In  \cite{GN_supint} we have shown that all classical superintegrable systems (and their generalizations not necessary superintegrable) in two-dimensional real
Euclidean space $E_2$ \cite{Fris1965} possess hidden symmetries leading to their linearization, as well as the Tremblay-Turbiner-Winternitz system \cite{TTW}, and a superintegrable system that is separable in cartesian coordinates and admits a third-order integral of motion as derived by Gravel in \cite{Gravel04}.
Then, we have conjectured that also
superintegrable systems in two-dimensional non-Euclidean space can be reduced to linear
equations by means of their hidden symmetries.
In this paper, we consider the two Perlick systems on two-dimensional non-Euclidean spaces
\cite{Perlick}, \cite{BalleQG08} \cite{Riglionietal_book13}, \cite{Riglioni13},
the two-dimensional Taub-NUT system \cite{Manton}, \cite{BalleSIGMA11},  \cite{Latini15},
and all the superintegrable systems for
 the four types of Darboux spaces as determined in \cite{KalKreWint}, \cite{KalKreMillWint}. We show that they are all intrinsically linear by determining their hidden Lie symmetries. As in \cite{harmony}, \cite{marcelnuc}, \cite{NucPost}, \cite{GN_supint}, \cite{PMNP15} we also make use of the reduction method \cite{kepler}.
More details on superintegrable systems and their hidden linearity have been described in \cite{GN_supint}. In particular, it is regardless of the separability of the corresponding Hamilton-Jacobi equation as shown in \cite{marcelnuc} for Kepler problem in cartesian coordinates, and in \cite{NucPost} for a superintegrable system in $E_2$ that does not allow
separation of variables \cite{PostWint11}.

\section{Perlick Type I}
We consider the so-called Hamiltonian of Perlik Type I \cite{Riglioni13}, i.e.:
\begin{equation}
H_{I}=\frac{(1+k r^2)^2}{2}\left(p_r^2+\frac{p_{\theta}^2}{r^2}\right)+A\frac{1-k r^2}{r}\label{HI}
\end{equation}
that generates the Hamiltonian equations:
\begin{equation} \left\{
\begin{array}{rcl}
\dot r&=&p_r(1+k r^2)^2,\\ [0.25cm]
\dot \theta &=&\displaystyle\frac{p_{\theta}(1+k r^2)^2}{r^2},\\ [0.35cm]
\dot p_r &=& \displaystyle \frac{\left(\left(1-kr^2\right)p_{\theta}^2-2kr^4p_r^2+A r\right)(1+k r^2)}{r^3},\\
\dot p_{\theta} &=& 0.
\end{array}  \right. \label{HIeq}
\end{equation}
The last equation can be easily integrated to give $p_{\theta}=w={\rm constant}$.
If we apply the reduction method developed in \cite{kepler} to the three remaining equations of system \eqref{HIeq}, and choose $\theta$ as a new independent
variable $y$, then we obtain the following two equations:
\begin{equation} \left\{
\begin{array}{rcl}
 \displaystyle\der{r}{y}&=&\displaystyle\frac{p_r r^2}{w},\\ [0.25cm]
\displaystyle\der{p_r}{y} &=&\displaystyle\frac{\left(1-kr^2\right)w^2-2kr^4p_r^2+A r}{rw(1+k r^2)},
\end{array}  \right. \label{HIeqred}
\end{equation}
If we derive $p_r$ from the first equation of system \eqref{HIeqred}  and replace it into the second equation, then we obtain the following second-order equation in $r$:
\begin{equation}
\displaystyle\der[2]{r}{y}=\frac{Ar^3 +(1- k r^2)w^2r^2 + 2w^2\left(\displaystyle\der{r}{y}\right)^2}{w^2r(1+kr^2)}.\label{HIeqo2}
\end{equation}
This equation admits an  eight-dimensional Lie symmetry algebra isomorphic to $\mathfrak{sl}(3,\R)$, and thus is linearizable. A two-dimensional abelian intransitive subalgebra is that generated by the two operators
\begin{equation}
\Gamma_7=\displaystyle\frac{\cos(y)r^2}{1+kr^2}\partial_r,\quad \Gamma_8=\displaystyle\frac{\sin(y)r^2}{1+kr^2}\partial_r
\end{equation}
that can be put into the canonical form \cite{Lie12} $\partial_{\tilde r}, \; \tilde y\partial_{\tilde r}$ by means of the transformation
\begin{equation}
\tilde y=\tan(y),\quad \tilde r=\displaystyle\frac{kr-r^{-1}-A/w^2}{\cos(y)}.
\end{equation}
Then equation \eqref{HIeqo2} becomes the free-particle equation $$\der[2]{\tilde r}{\tilde y}=0.$$
Instead if we make the transformation of the dependent variable $u=kr-r^{-1}-A/w^2$ only, then equation \eqref{HIeqo2} becomes the equation of the harmonic oscillator $$\displaystyle\der[2]{u}{y}=-u.$$

\section{Perlick Type II}
We consider the so-called Hamiltonian of Perlik Type II \cite{Riglioni13}, i.e.:
\begin{equation}
H_{II}=\frac{(1-\lambda^2 r^4)^2}{2(1+\lambda^2r^4-2\delta r^2)}\left(p_r^2+\frac{p_{\theta}^2}{r^2}\right)+\frac{B r^2}{1+\lambda^2r^4-2\delta r^2}\label{HII}
\end{equation}
that generates the Hamiltonian equations:
\begin{equation} \left\{
\begin{array}{rcl}
\dot r&=&\displaystyle\frac{p_r(1-\lambda^2 r^4)^2}{1+\lambda^2r^4-2\delta r^2},\\ [0.25cm]
\dot \theta &=&\displaystyle\frac{p_{\theta}(1-\lambda^2 r^4)^2}{r^2(1+\lambda^2r^4-2\delta r^2)},\\ [0.35cm]
\dot p_r &=& \displaystyle \frac{1-\lambda^2 r^4}{r^3(1+\lambda^2r^4-2\delta r^2)^2}\left(2r^2p_r^2\left(\lambda^4 r^6 +  3 \lambda^2 r^2 -\delta-3\delta \lambda^2 r^4\right)
\right.\\ [0.35cm] && \left. \quad+ p_{\theta}^2 \left(1+ \lambda^4 r^8 +6 \lambda^2 r^4  - 4 \delta \lambda^2 r^6  -4  \delta r^2\right)- 2 B r^4\right),\\
\dot p_{\theta} &=& 0.
\end{array}  \right. \label{HIIeq}
\end{equation}
The last equation can be easily integrated to give $p_{\theta}=w={\rm constant}$.
If we apply the reduction method developed in \cite{kepler} to the three remaining equations of system \eqref{HIIeq}, and choose $\theta$ as a new independent
variable $y$, then we obtain the following two equations:
\begin{equation} \left\{
\begin{array}{rcl}
 \displaystyle\der{r}{y}&=&\displaystyle\frac{p_r r^2}{w},\\ [0.35cm]
\displaystyle\der{p_r}{y} &=&\displaystyle\frac{1}{w r(1+\lambda^2r^4-2\delta r^2)(1-\lambda^2 r^4)}\left(2r^2p_r^2\left(\lambda^4 r^6 +  3 \lambda^2 r^2 -\delta-3\delta \lambda^2 r^4\right)
\right.\\ [0.35cm] && \left. \quad+ w^2 \left(1+ \lambda^4 r^8 +6 \lambda^2 r^4  - 4 \delta \lambda^2 r^6  -4  \delta r^2\right)- 2 B r^4\right),
\end{array}  \right. \label{HIIeqred}
\end{equation}
If we derive $p_r$ from the first equation of system \eqref{HIIeqred}  and replace it into the second equation, then we obtain the following second-order equation in $r$:
\begin{equation}
\displaystyle\der[2]{r}{y}=\displaystyle\frac{w^2\left(2\left(\der{r}{y}\right)^2\left(\lambda^4 r^6 +  3 \lambda^2 r^2 -\delta-3\delta \lambda^2 r^4\right)+ 1+ \lambda^4 r^8 +6 \lambda^2 r^4  - 4 \delta \lambda^2 r^6  -4  \delta r^2\right)- 2 B r^4}{w r(1+\lambda^2r^4-2\delta r^2)(1-\lambda^2 r^4)},\label{HIIeqo2}
\end{equation}
that admits a three-dimensional symmetry algebra $\Sl(2,\R)$, unless $B=2w^2( \lambda^2- \delta^2)$ in which case it admits an eight-dimensional Lie symmetry algebra $\Sl(3,\R)$ and thus it is linearizable. We now use the general method described in \cite{Leach2003} and that may be
applied to any second-order ordinary differential equation that admits a Lie
symmetry algebra $sl(2,\R)$. If we solve equation \eqref{HIIeqo2} with respect to
$B$ and derive once with respect to $y$, then we obtain the following third-order equation:
\begin{equation}
\displaystyle\der[3]{r}{y}=\displaystyle\frac{\displaystyle\der{r}{y}}{r^2(1-\lambda^2 r^4)}\left(3r(3+\lambda^2r^4)\displaystyle\der[2]{r}{y}- 12\left(\der{r}{y}\right)^2
-4r^2(1-\lambda^2r^4) \right). \label{HIIeqo3}
\end{equation}
which admits a seven-dimensional Lie symmetry algebra, and therefore is linearizable. Indeed, the new dependent variable $\tilde r=\frac{1+\lambda^2r^4}{2r^2}$ transforms equation \eqref{HIIeqo3} into the linear equation
$$\displaystyle\der[3]{\tilde r}{y}=-4\displaystyle\der{\tilde r}{y},$$
which is once-derived linear harmonic oscillator with frequency equal to 2.

\subsection{Taub-NUT}

The following Taub-NUT Hamiltonian \cite{BalleSIGMA11}, \cite{Latini15}:
\begin{equation}
    H_{\text{TN}}(\eta) =
    \half\frac{r}{\eta + r}
    \left( p_{r}^{2} + \frac{1}{r^{2}} p_{\varphi}^{2} \right)
    - \frac{\alpha}{\eta+r}
    \label{eqn:hamtaubnut}
\end{equation}
yields the Hamiltonian equations:
\begin{equation} \left\{
\begin{array}{rcl}
        \dot{r} &=& \displaystyle\frac{r p_r}{\eta+r},
           \\ [0.25cm]
        \dot{\varphi}&=&\displaystyle\frac{p_{\varphi}}{\left(\eta+r\right) r},
               \\ [0.25cm]
        \dot p_{r}&=&-\displaystyle
        \frac{2 \left(\alpha r-p_{\varphi}^{2}\right) r
        +\eta\left(r^{2} p_r^{2} - p_{\varphi}^{2}\right)}%
        {2 \left(\eta+r\right)^{2} r^{2}},
              \\ [0.25cm]
        \dot{p_{\varphi}} &=& 0.
\end{array}  \right. \label{eqn:hameqshamtaubnutmod}
\end{equation}
The last equation can be easily integrated to give $p_{\varphi}=w_{0}={\rm constant}$.
If we apply the reduction method developed in \cite{kepler} to the three remaining equations of system \eqref{eqn:hameqshamtaubnutmod}, and choose $\varphi$ as a new independent
variable $y$, then we obtain two equations:
\begin{equation} \left\{
\begin{array}{rcl}
  \displaystyle \der{r}{y} &=& \displaystyle\frac{r^{2}}{w_{0}}p_r,\\ [0.25cm]
   \displaystyle \der{p_r}{y} &=&\displaystyle \frac{2 \left(\alpha r-w_0^{2}\right) r
        +\eta\left(r^{2} p_r^{2} - w_0^{2}\right)}{2 \left(\eta+r\right) rw_0}.
  \end{array}  \right.  \label{eqn:hameqhamtaubnutmoda2}
\end{equation}
Solving the first equation for $p_r$ and substituting into the second
 yields:
\begin{equation}
    \der[2]{u}{y}=
    \frac{3 \eta+4 u}{2 u \left(\eta+u\right)}
    \left( \der{u}{y} \right)^{2}
    -\frac{u \left(2 \alpha u^{2}-\eta w_{0}^{2}-2w_{0}^{2} u \right)}%
    {2 w_{0}^{2} \left(\eta+u\right)},
    \label{eqn:hameqhamtaubnutmodc2}
\end{equation}
with $u\equiv r$.
This equation \eqref{eqn:hameqhamtaubnutmodc2}
admits a three-dimensional Lie symmetry algebra  spanned by the following operators:
\begin{equation}
    \begin{gathered}
        \Theta_{1} = \partial_{y},
        \quad
        \Theta_{2} = \cos (y) \partial_{y} +
        \frac{u(\eta+u)}{\eta}\sin (y) \partial_{u},
        \\
        \Theta_{3} = \sin (y) \partial_{y} -
        \frac{u(\eta+u)}{\eta}\cos(y) \partial_{u},
    \end{gathered}
    \label{eqn:symmhameqhamtaubnutmodc2}
\end{equation}
However if  $\alpha=0$ then the equation admits an eight-dimensional Lie symmetry algebra.
Therefore if we solve equation \eqref{eqn:hameqhamtaubnutmodc2} with respect to $\alpha$ and derive once with respect to $y$, then we get the following third-order equation:
\begin{equation}
    u^{2}\der[3]{u}{y}
    + \der{u}{y} \left[u^{2} - 6 u \der[2]{u}{y} + 6 \left(\der{u}{y}\right)^{2}\right] = 0,
    \label{eqn:raisedorderhameqshamtaubnutmodd}
\end{equation}
which is linearizable since it admits a seven-dimensional Lie symmetry algebra spanned by the following operators:
\begin{equation}
    \begin{gathered}
        \Pi_{1} =\partial_y,\quad \Pi_{2} =\cos (y) \partial_y+u\sin (y) \partial_u, \quad \Pi_{3} =\sin (y) \partial_y-u\cos (y) \partial_u,
        \\
        \Pi_{4}=u\partial_u,\quad \Pi_{5} ={u}^{2}\partial_u,\quad \Pi_{6}={u}^{2}\cos (y) \partial_u,\quad \Pi_{7} ={u}^{2}\sin (y) \partial_u.
   \end{gathered}
    \label{eqn:symmraisedordertaubnut}
\end{equation}
A two-dimensional Abelian intransitive subalgebra is that generated by the operators  $\Pi_{6}$ and $\Pi_{7}$. If we put them into the canonical form $\partial_U, Y\partial_U$, then the transformation
\begin{equation}
    Y = \frac{\sin(y)}{\cos(y)}, \quad U = - \frac{1}{u \cos (y)}
    \label{eqn:linearizing1raisedordertaubnut}
\end{equation}
changes equation \eqref{eqn:hameqhamtaubnutmodc2} into
the following linear equation:
\begin{equation}
    \der[3]{U}{Y}=-3 \frac{Y^{2}}{1+Y^{2}}\der[2]{U}{Y}.
    \label{eqn:linearization1raisedordertaubnut}
\end{equation}
Moreover, if  we consider the  transformation  $v = - \frac{1}{u}$, then
equation \eqref{eqn:hameqhamtaubnutmodc2} becomes the once-derived linear harmonic oscillator with frequency equal to 1, i.e.:
\begin{equation}
    \der[3]{v}{y}=- \der{v}{y}.
    \label{eqn:linearization2raisedordertaubnut}
\end{equation}
This shows the connection between the Taub-NUT
Hamiltonian \eqref{eqn:hamtaubnut} and the harmonic oscillator.

\section{Darboux I}
Three superintegrable systems were determined in
\cite{KalKreWint}, where the problem of superintegrability for the Hamiltonian
\begin{equation}
H_{DI}=\frac{1}{4u}\left(p_u^2+p_v^2\right)+V(u,v)
\end{equation}
was addressed, namely finding the potentials $V(u,v)$ such that $H_{DI}$ admits at
least two extra quadratic integrals. We show that all of three systems have hidden symmetries that make them linear.
\subsection{Case (1)}
The
Hamiltonian
\begin{equation}
{\cal H}_{DI1}=\frac{1}{4u}\left(p_u^2+p_v^2\right)+b_1\frac{4u^2+v^2}{4u}+\frac{b_2}{u}+\frac{b_3}{uv^2}
\end{equation}
yields the Hamiltonian equations
\begin{equation} \left\{
\begin{array}{rcl}
\dot u&=&\displaystyle \frac{p_u}{2u},\\ [0.25cm]
\dot v&=& \displaystyle \frac{p_v}{2u},\\ [0.25cm]
\dot p_u &=& \displaystyle{\frac{v^2(p_u^2+p_v^2)+b_1v^2(v^2-4u^2)+4b_2v^2+4b_3}{4u^2v^2}},\\[0.25cm]
\dot p_v &=&\displaystyle{\frac{4b_3-b_1v^4}{2uv^3}}.
\end{array}  \right. \label{HDI1eq}
\end{equation}
If we apply the reduction method developed in \cite{kepler}  and choose $v$ as a new independent
variable $y$, then we obtain the following three equations:
\begin{equation} \left\{
\begin{array}{rcl}
 \displaystyle\der{u}{y}&=&\displaystyle\frac{p_u}{p_v},\\ [0.35cm]
\displaystyle\der{p_u}{y} &=&\displaystyle{\frac{y^2(p_u^2+p_v^2)+b_1y^2(y^2-4u^2)+4b_2y^2+4b_3}{2uy^2p_v}},\\ [0.35cm]
\displaystyle\der{p_v}{y} &=&\displaystyle{\frac{4b_3-b_1y^4}{y^3p_v}},
\end{array}  \right. \label{HDI1eqred}
\end{equation}
The last equation can be easily integrated, i.e.:
\begin{equation}
p_v=\pm\displaystyle \frac{\sqrt{2 w_0 b_1y^2 - b_1y^4 - 8w_0b_3y^2 - 4b_3}}{y},
\end{equation}
with $w_0$ an arbitrary constant. Moreover, if we derive $p_u$ from the first equation of system \eqref{HDI1eqred}  and replace it into the second equation, then we obtain the following second-order equation in $u$:
\begin{equation}
\displaystyle\der[2]{u}{y}=\frac{1}{2u}\left(\displaystyle\der{u}{y}\right)^2+\frac{u(b_1y^4-b_3)\displaystyle\der{u}{y}+y^3(w_0b_1-2b_1u^2+2b_2-4w_0b_3)}{yu(2 w_0 b_1y^2 - b_1y^4 - 8w_0b_3y^2 - 4b_3)},\label{HDI1eqo2}
\end{equation}
that admits a three-dimensional symmetry algebra $\Sl(2,\R)$, unless $b_2=0$ in which case it admits an eight-dimensional Lie symmetry algebra $\Sl(3,\R)$ and thus it is linearizable. We now use the general method described in \cite{Leach2003} and that may be
applied to any second-order ordinary differential equation that admits a Lie
symmetry algebra $sl(2,\R)$. If we solve equation \eqref{HDI1eqo2} with respect to
$b_2$ and derive once with respect to $y$, then we obtain the following linear third-order equation:
\begin{equation}
\displaystyle\der[3]{u}{y}=\displaystyle\frac{(b_1y^4-4b_3)\left(\displaystyle\der{u}{y}-y\displaystyle\der[2]{u}{y}\right)}{y^2(2 w_0 b_1y^2 - b_1y^4 - 8w_0b_3y^2 - 4b_3)}.\label{HDI1eqo3}
\end{equation}

\subsection{Case (2)}
The
Hamiltonian
\begin{equation}
{\cal H}_{DI2}=\frac{1}{4u}\left(p_u^2+p_v^2\right)+\frac{a_1}{u}+\frac{a_2v}{u}+a_3\frac{u^2+v^2}{u}
\end{equation}
yields the Hamiltonian equations
\begin{equation} \left\{
\begin{array}{rcl}
\dot u&=&\displaystyle \frac{p_u}{2u},\\ [0.25cm]
\dot v&=& \displaystyle \frac{p_v}{2u},\\ [0.25cm]
\dot p_u &=& \displaystyle{\frac{p_u^2+ p_v^2+4a_1+4a_2v^2-4a_3(u^2-v^2)}{4u^2}},\\[0.25cm]
\dot p_v &=&-\displaystyle{\frac{a_2+2a_3v}{u}}.
\end{array}  \right. \label{HDI2eq}
\end{equation}
If we apply the reduction method developed in \cite{kepler}  and choose $v$ as a new independent
variable $y$, then we obtain the following three equations:
\begin{equation} \left\{
\begin{array}{rcl}
 \displaystyle\der{u}{y}&=&\displaystyle\frac{p_u}{p_v},\\ [0.35cm]
\displaystyle\der{p_u}{y} &=&\displaystyle{\frac{4a_1+4a_2y-4a_3u^2+4a_3y^2+p_u^2+p_v^2}{2up_v}},\\ [0.35cm]
\displaystyle\der{p_v}{y} &=&-2\displaystyle{\frac{a_2+2a_3y}{p_v}},
\end{array}  \right. \label{HDI2eqred}
\end{equation}
The last equation can be easily integrated, i.e.:
\begin{equation}
p_v=\pm\displaystyle 2\sqrt{a_2w_0-a_2y-a_3y^2},
\end{equation}
with $w_0$ an arbitrary constant. Moreover, if we derive $p_u$ from the first equation of system \eqref{HDI2eqred}  and substitute it into the second equation, then we obtain the following second-order equation in $u$:
\begin{equation}
\displaystyle\der[2]{u}{y}=\displaystyle\frac{(a_2w_0-a_2y-a_3y^2)\displaystyle\left(\der{u}{y}\right)^2+(a_2+2a_3y)u\displaystyle\der{u}{y}-a_3u^2+a_1+a_2w_0}{2u(a_2w_0-a_2y-a_3y^2)},\label{HDI2eqo2}
\end{equation}
that admits a three-dimensional symmetry algebra $\Sl(2,\R)$, unless $a_1+a_2 w_0=0$ in which case it admits an eight-dimensional Lie symmetry algebra $\Sl(3,\R)$ and thus it is linearizable. We now use the general method described in \cite{Leach2003} and that may be
applied to any second-order ordinary differential equation that admits a Lie
symmetry algebra $\Sl(2,\R)$. If we solve equation \eqref{HDI2eqo2} with respect to
$a_1$ and derive once with respect to $y$, then we obtain the following linear third-order equation:
\begin{equation}
\displaystyle\der[3]{u}{y}=\displaystyle\frac{3(a_2+2a_3y)}{2(a_2w_0-a_2y-a_3y^2)}\der[2]{u}{y}.\label{HDI2eqo3}
\end{equation}

\subsection{Case (3)}
 The
Hamiltonian
\begin{equation}
{\cal H}_{DI3}=\frac{1}{4u}\left(p_u^2+p_v^2\right)+\frac{a}{u}
\end{equation}
yields the Hamiltonian equations
\begin{equation} \left\{
\begin{array}{rcl}
\dot u&=&\displaystyle\frac{p_u}{2u},\\ [0.25cm]
\dot v&=& \displaystyle\frac{p_v}{2u},\\ [0.25cm]
\dot p_u &=& \displaystyle{\frac{4a+p_u^2+ p_v^2}{4u^2}},\\[0.25cm]
\dot p_v &=&0.
\end{array}  \right. \label{HDI3eq}
\end{equation}
The last equation can be easily integrated, i.e. $p_v=w_0$, with $w$ an arbitrary constant. If we apply the reduction method developed in \cite{kepler}  and  choose $v$ as new independent
variable, then system \eqref{HDI3eq} reduces to the following two equations:
\begin{equation} \left\{
\begin{array}{ccl}
\displaystyle\der{u}{y}&=&\displaystyle\frac{p_u}{w_0},\\ [0.35cm]
\displaystyle\der{p_u}{y} &=&\displaystyle{\frac{4a+p_u^2+ w_0^2}{2uw_0}}.\label{HDI3eqred}
\end{array}  \right.
\end{equation}
If we derive $p_u$ from the first equation of system \eqref{HDI3eqred}  and replace it into the second equation, then we obtain the following second-order equation in $u$:
\begin{equation}
\frac{{\rm d}^2u}{{\rm d}y^2}=\frac{1}{2u}\left(\frac{{\rm d}u}{{\rm d}v}\right)^2+\frac{4a+w_0^2}{2w_0^2u},\label{HDI3eqo2}
\end{equation}
that admits a three-dimensional symmetry algebra $\Sl(2,\R)$, unless $4a+ w_0^2=0$ in which case it admits an eight-dimensional Lie symmetry algebra $\Sl(3,\R)$ and thus it is linearizable. We now use the general method described in \cite{Leach2003} and that may be
applied to any second-order ordinary differential equation that admits a Lie
symmetry algebra $\Sl(2,\R)$. If we solve equation \eqref{HDI3eqo2} with respect to
$a$ and derive once with respect to $y$, then we obtain the following linear third-order equation:
\begin{equation}
\frac{{\rm d}^3u}{{\rm d}y^3}=0.
\end{equation}

\section{Darboux II}
Four superintegrable systems were determined in
\cite{KalKreMillWint}, where the problem of superintegrability for the Hamiltonian
\begin{equation}
H_{DII}=\frac{w_1^2}{w_1^2+1}\left(w_3^2+w_4^2\right)+V(w_1,w_2)
\end{equation}
was addressed, namely finding the potentials $V(w_1,w_2)$ such that $H_{DII}$ admits at
least two extra quadratic integrals.
\subsection{Case (A)}
 The
Hamiltonian
\begin{equation}
{\cal H}_{DIIA}=\frac{w_1^2}{w_1^2+1}\left(w_3^2+w_4^2+a_1\left(\frac{w_1^2}{4}+w_2^2\right)+a_2w_2+\frac{a_3}{w_1^2}\right)
\end{equation}
yields the Hamiltonian equations
\begin{equation} \left\{
\begin{array}{rcl}
\dot w_1&=&2\displaystyle\frac{w_1^2w_3}{w_1^2+1},\\ [0.3cm]
\dot w_2&=& 2\displaystyle\frac{w_1^2w_4}{w_1^2+1},\\ [0.3cm]
\dot w_3 &=& -w_1\displaystyle{\frac{4w_3^2 + 4w_4^2+a_1 w_1^4+2 a_1 w_1^2 + 4 a_1 w_2^2 + 4 a_2 w_2 - 4 a_3 }{2(w_1^2 + 1)^2}},\\[0.3cm]
\dot w_4 &=& -w_1^2\displaystyle{\frac{2a_1w_2+a_2}{w_1^2 + 1}}.
\end{array}  \right. \label{HDIIAeq}
\end{equation}
If we apply the reduction method developed in \cite{kepler}  and choose $w_1$ as a new independent
variable $y$, then we obtain the following three equations:
\begin{equation} \left\{
\begin{array}{rcl}
 \displaystyle\der{w_2}{y}&=&\displaystyle\frac{w_4}{w_3},\\ [0.35cm]
\displaystyle\der{w_3}{y} &=&-\displaystyle{\frac{4w_3^2 + 4w_4^2+a_1 y^4+2 a_1 y^2 + 4 a_1 w_2^2 + 4 a_2 w_2 - 4 a_3}{2yw_3(w_1^2 + 1)}},\\ [0.35cm]
\displaystyle\der{w_4}{y} &=&-\displaystyle{\frac{2a_1w_2+a_2}{w_3}}.
\end{array}  \right. \label{HDIIAeqred}
\end{equation}
If we solve the second equation with respect to $a_3$ and then derive once with respect to y, then we obtain the following second-order equation in $w_3(y)$:
\begin{equation}
w_3''=-\displaystyle{\frac{w_3'\left(yw_3'+3w_3\right)+a_1y}{yw_3}},\label{HDIIAeqo2}
\end{equation}
that admits an eight-dimensional Lie symmetry algebra $\Sl(3,\R)$ and therefore it is linearizable. In this case, Lie canonical transformation is such:
\begin{equation}\tilde w_3 =\displaystyle{\frac{y^2w_3}{2}+\frac{a_1y^4}{8}}, \quad \tilde y=y^2 \Longrightarrow \displaystyle \frac{{\rm d}^2\tilde w_3}{{\rm d} \tilde y^2}=0,\end{equation}
and consequently
\begin{equation}
w_3=\pm \displaystyle{\frac{\sqrt{8C_2y^2 + 8C_1 - a_1y^4}}{2y}},
\end{equation}
with $C_1,C_2$ arbitrary integration constants, although only one is really arbitrary since there is a relationship between them and $a_3$. Then, if we solve the first equation in \eqref{HDIIAeqred} with respect to $w_4$, and replace it into the third equation, we obtain the following linear second-order equation in $w_2(y)$:
 \begin{equation}
 w_2''=\displaystyle{ \frac{-(a_1y^4+8C_1)w_2' + 4a_1y^3w_2 + 2a_2y^3}{a_1y^5 - 8C_1 y - 8C_2y^3}},
 \end{equation}
 and its general solution is
 \begin{equation}w_2=(a_1 y^2-4C_2) C_3+\sqrt{a1 y^4-8 C_1-8 C_2 y^2} C_4-\frac{a_2}{2a_1},     \end{equation}
 with $C_3,C_4$ arbitrary integration constants.

\subsection{Case (B)}
 The
Hamiltonian
\begin{equation}
{\cal H}_{DIIB}=\frac{w_1^2}{w_1^2+1}\left(w_3^2+w_4^2+b_1(w_1^2+w_2^2)+\frac{b_2}{w_1^2}+\frac{b_3}{w_2^2}\right)
\end{equation}
yields the Hamiltonian equations
\begin{equation} \left\{
\begin{array}{rcl}
\dot w_1&=&2\displaystyle\frac{w_1^2w_3}{w_1^2+1},\\ [0.3cm]
\dot w_2&=& 2\displaystyle\frac{w_1^2w_4}{w_1^2+1},\\ [0.3cm]
\dot w_3 &=& -2w_1\displaystyle{\frac{w_2^2w_3^2+w_2^2w_4^2 +b_1w_1^4 w_2^2 + 2 b_1w_1^2w_2^2 + b_1 w_2^4 - b_2w_2^2 + b_3  }{w_2^2(w_1^2 + 1)^2}},\\[0.3cm]
\dot w_4 &=& -2w_1^2\displaystyle{\frac{b_1w_2^4-b_3}{w_2^3(w_1^2 + 1)}}.
\end{array}  \right. \label{HDIIBeq}
\end{equation}
If we apply the reduction method developed in \cite{kepler}  and choose $w_1$ as a new independent
variable $y$, then we obtain the following three equations:
\begin{equation} \left\{
\begin{array}{rcl}
 \displaystyle\der{w_2}{y}&=&\displaystyle\frac{w_4}{w_3},\\ [0.35cm]
\displaystyle\der{w_3}{y} &=&-\displaystyle{\frac{w_2^2w_3^2+w_2^2w_4^2 +b_1y^4 w_2^2 + 2 b_1y^2w_2^2 + b_1 w_2^4 - b_2w_2^2 + b_3 }{w_2^2w_3y(y^2 + 1)}},\\ [0.35cm]
\displaystyle\der{w_4}{y} &=&-\displaystyle{\frac{b_1w_2^4-b_3}{w_2^3w_3}}.
\end{array}  \right. \label{HDIIBeqred}
\end{equation}
If we solve the second equation with respect to $b_2$ and then derive once with respect to y, then we obtain the following second-order equation in $w_3(y)$:
\begin{equation}
w_3''=-\displaystyle{\frac{w_3'\left(yw_3'+3w_3\right)+4b_1y}{yw_3}},\label{HDIIBeqo2}
\end{equation}
which is exactly the linearizable equation \eqref{HDIIAeqo2} if  $a_1$ is replaced with $4b_1$, and consequently
\begin{equation}
w_3=\pm \displaystyle{\frac{\sqrt{8C_2y^2 + 8C_1 - 4b_1y^4}}{2y}},
\end{equation}
with $C_1,C_2$ arbitrary integration constants. \\
Another reduction would also lead to linearity. If we choose $w_2$ as a new independent
variable $y$, then we obtain the following three equations:
\begin{equation} \left\{
\begin{array}{rcl}
 \displaystyle\der{w_1}{y}&=&\displaystyle\frac{w_3}{w_4},\\ [0.35cm]
\displaystyle\der{w_3}{y} &=&-\displaystyle{\frac{y^2w_3^2+y^2w_4^2 +b_1w_1^4 y^2 + 2 b_1w_1^2y^2 + b_1 y^4 - b_2y^2 + b_3 }{w_1w_4y^2(w_1^2 + 1)}},\\ [0.35cm]
\displaystyle\der{w_4}{y} &=&-\displaystyle{\frac{b_1y^4-b_3}{y^3w_4}},
\end{array}  \right. \label{HDIIBeqred2}
\end{equation}
and we can easily integrate the third equation, i.e.
\begin{equation}
w_4=\pm\frac{1}{y}\sqrt{2b_1w_0y^2-b_1y^4-2b_3w_0y^2-b_3},
\end{equation}
with $w_0$ an arbitrary integration constant.
Then, if we solve the first equation in \eqref{HDIIBeqred2} with respect to $w_3$, and replace it into the second equation, we obtain the following second-order equation in $w_1(y)$:
\begin{equation}
w_1''=\displaystyle \frac{-w_1'^2}{w_1(w_1^2+1)}+\frac{(b_1y^4 - b_3)w_1'}{y(2(b_1-b_3)w_0y^2 - b_1y^4-b3)}-y^2\frac{b_1(w_1^4+2w_1^2+2w_0)-b_2-2b_3w_0}{w_1(w_1^2+1)(2(b_1-b_3)w_0y^2 - b_1y^4-b_3)}. \label{HIIBeqo2b}
\end{equation}
that admits a three-dimensional symmetry algebra $\Sl(2,\R)$, unless $b_2=2(b_1-b_3)w_0-b_1$, in which case it admits an eight-dimensional Lie symmetry algebra $\Sl(3,\R)$ and thus it is linearizable. We now use the general method described in \cite{Leach2003} and that may be
applied to any second-order ordinary differential equation that admits a Lie
symmetry algebra $\Sl(2,\R)$. If we solve equation \eqref{HIIBeqo2b} with respect to
$b_2$ and derive once with respect to $y$, then we obtain the following third-order equation:
\begin{equation}
w_1'''=-3\displaystyle\frac{w_1'w_1''}{w_1}+3\frac{b_1y^4 - b_3}{2(b_1-b_3)w_0y^2 - b_1y^4-b_3}\left(\displaystyle\frac{w_1''}{y}+\frac{w_1'^2}{yw_1}-\frac{w_1'}{y^2} \right), \label{HIIBeqo3}
\end{equation}
which admits a seven-dimensional Lie symmetry algebra, and therefore is linearizable. Indeed, the new dependent variable $U=\frac{1}{2w_1^2}$ and independent variable
$Y=b_1y^2+w_0(b_3-b_1)$ transform equation \eqref{HIIBeqo3} into the linear equation
$$\displaystyle\der[3]{U}{Y}=-\displaystyle\frac{3Y\displaystyle\der[2]{U}{Y}}{Y^2-b_3^2w_0^2+b_1b_3+2b_1b_3w_0^2-b_1^2w_0^2}.$$

\subsection{Case (C)}
 The
Hamiltonian
\begin{equation}
{\cal H}_{DIIC}=\displaystyle\frac{w_3^2+w_4^2+a_1+\frac{a_2}{w_1^2}+\frac{a_3}{w_2^2}}{w_1^2+w_2^2+\frac{1}{w_1^2}+\frac{1}{w_2^2}}
\end{equation}
yields the Hamiltonian equations
\begin{equation} \left\{
\begin{array}{rcl}
\dot w_1&=&2\displaystyle\frac{w_1^2w_2^2w_3}{(w_1^2w_2^2 + 1)(w_1^2 + w_2^2)},\\ [0.3cm]
\dot w_2&=& 2\displaystyle\frac{w_1^2w_2^2w_4}{(w_1^2w_2^2 + 1)(w_1^2 + w_2^2)},\\ [0.3cm]
\dot w_3 &=& 2w_1w_2^2\displaystyle{\frac{(a_1w_2^2 + a_3 + (w_3^2 + w_4^2)w_2^2)(w_1^4- 1) + (w_2^4 + 1 + 2w_1^2w_2^2)a_2}{(w_1^2w_2^2 + 1)^2(w_1^2 + w_2^2)^2}},\\[0.3cm]
\dot w_4 &=& 2w_1^2w_2\displaystyle{\frac{(a_1w_1^2 + a_2 + (w_3^2 + w_4^2)w_1^2)(w_2^4- 1) + (w_1^4 + 2w_1^2w_2^2 + 1)a_3}{(w_1^2w_2^2 + 1)^2(w_1^2 + w_2^2)^2}}.
\end{array}  \right. \label{HDIICeq}
\end{equation}
Before applying the reduction method  \cite{kepler}, we introduce the following transformations of the dependent variables in order to avoid the mishandling of formulas with square roots by either REDUCE or MAPLE, i.e.
\begin{equation}
w_1=\sqrt{r_1},\quad w_2=\sqrt{r_2},\quad w_3=\sqrt{r_3},\quad w_4=\sqrt{r_4}, \label{transfwisqrtri}
\end{equation}
and then choose  $r_2$ as a new independent variable $y$ which gives rise to the following three equations:
\begin{equation} \left\{
\begin{array}{rcl}
 \displaystyle\der{r_1}{y}&=&\displaystyle\sqrt{\frac{r_1r_3}{yr_4}},\\ [0.35cm]
\displaystyle\der{r_3}{y} &=&\displaystyle{\sqrt{\frac{r_3}{yr_1r_4}}\frac{(a_1y + a_3 + (r_3+ r_4)y)(r_1^2- 1) + (y^2 + 1 + 2r_1y)a_2}{(r_1y + 1)(r_1 + y)}},\\ [0.35cm]
\displaystyle\der{r_4}{y} &=&\displaystyle{\frac{(a_1r_1 + a_2 + (r_3+ r_4)r_1)(y^2- 1) + (r_1^2 + 1 + 2r_1y)a_3}{y(r_1y + 1)(r_1 + y)}}.
\end{array}  \right. \label{HDIICeqred}
\end{equation}
From the Hamiltonian ${\cal H}_{DIIC}$, i.e.
\begin{equation}
{\cal H}_{DIIC}=\displaystyle\frac{r_3+r_4 +a_1+\frac{a_2}{r_1 }+\frac{a_3}{y }}{r_1 +y +\frac{1}{r_1 }+\frac{1}{y }}=h_0,
\end{equation}
we can derive:
\begin{equation}
r_3=\displaystyle\frac{(r_1y + 1)(r_1 + y)h_0-r_1r_4y-a_3r_1-a_2y-a_1yr_1}{yr_1},
\end{equation}
with $h_0$ an arbitrary constant. Consequently, the third equation in \eqref{HDIICeqred} becomes:
\begin{equation}
\displaystyle\der{r_4}{y}=\displaystyle\frac{a_3+(y^2-1)h_0}{y^2},
\end{equation}
 that can be easily integrated, i.e.:
 \begin{equation}
r_4=\displaystyle\frac{w_0y-a_3+(y^2+1)h_0}{y^2},
\end{equation}
 with $w_0$ an arbitrary constant. Finally,  we are left with the first equation in \eqref{HDIICeqred}, i.e.
 \begin{equation}
 \displaystyle\der{r_1}{y}=\displaystyle\frac{\sqrt{h_0r_1^2-(a_1+w_0)r_1-a_2+h_0}}{\sqrt{h_0y^2+w_0y-a_3+h_0}},
 \end{equation}
 which can be solved by quadratures. However, if we solve it with respect to $a_2$, and derive once by $y$, then the following linear second-order equation is obtained:
 \begin{equation}
 2\left(a_3-w_0y-(y^2+1)h_0\right)\displaystyle\der[2]{r_1}{y}-(w_0+2h_0y)\displaystyle\der{r_1}{y}+2h_0r_1-w_0-a_1=0.
 \end{equation}

\subsection{Case (D)}
 The
Hamiltonian
\begin{equation}
{\cal H}_{DIID}=\frac{w_1^2}{w_1^2+1}\left(w_3^2+w_4^2 +d \right)
\end{equation}
yields the Hamiltonian equations
\begin{equation} \left\{
\begin{array}{rcl}
\dot w_1&=&2\displaystyle\frac{w_1^2w_3}{w_1^2+ 1},\\ [0.3cm]
\dot w_2&=& 2\displaystyle\frac{w_1^2w_4}{w_1^2+ 1},\\ [0.3cm]
\dot w_3 &=& -2w_1\displaystyle{\frac{w_3^2+w_4^2 +d}{(w_1^2+ 1)^2}},\\[0.3cm]
\dot w_4 &=& 0.
\end{array}  \right. \label{HDIIDeq}
\end{equation}
The last equation yields $w_4=w_0$. If we apply the reduction method developed in \cite{kepler}  and choose $w_2$ as a new independent
variable $y$, then we obtain the following two equations:
\begin{equation} \left\{
\begin{array}{rcl}
 \displaystyle\der{w_1}{y}&=&\displaystyle\frac{w_3}{w_0},\\ [0.35cm]
\displaystyle\der{w_3}{y} &=&-\displaystyle{\frac{w_3^2+w_0^2 +d}{w_0w_1(w_1^2 + 1)}}.
\end{array}  \right. \label{HDIIDeqred}
\end{equation}
Then, if we solve the first equation in \eqref{HDIIDeqred} with respect to $w_3$, and replace it into the second equation, we obtain the following second-order equation in $w_1(y)$:
\begin{equation}
w_1''=-\displaystyle \frac{w_0^2w_1'^2 + w_0^2+ d}{w_0^2w_1(w_1^2 + 1)}. \label{HIIDeqo2}
\end{equation}
that admits a three-dimensional symmetry algebra $\Sl(2,\R)$, unless $d=-w_0^2$, in which case it admits an eight-dimensional Lie symmetry algebra $\Sl(3,\R)$ and thus it is linearizable. We now use the general method described in \cite{Leach2003} and that may be
applied to any second-order ordinary differential equation that admits a Lie
symmetry algebra $\Sl(2,\R)$. If we solve equation \eqref{HIIDeqo2} with respect to
$d$ and derive once with respect to $y$, then we obtain the following third-order equation:
\begin{equation}
w_1'''=-\displaystyle\frac{3w_1'w_1''}{w_1}, \label{HIIDeqo3}
\end{equation}
which admits a seven-dimensional Lie symmetry algebra, and therefore is linearizable. Indeed, the new dependent variable $r_1=w_1^2$ transforms equation \eqref{HIIDeqo3} into the linear equation
\begin{equation}r_1'''=0.\end{equation}

\section{Darboux III}
Five superintegrable cases were determined in
\cite{KalKreMillWint}, where the problem of superintegrability for the Hamiltonian
\begin{equation}
H_{DIII}=\frac{e^{2u}\left(p_u^2+p_v^2\right)}{4e^{u+1}}
\end{equation}
was addressed.
\subsection{Case (A)}
 The
Hamiltonian
\begin{equation}
{\cal H}_{DIIIA}=\displaystyle\frac{w_3^2+w_4^2+a_1w_1+a_2w_2+a_3}{4+w_1^2+w_2^2}
\end{equation}
yields the Hamiltonian equations
\begin{equation} \left\{
\begin{array}{rcl}
\dot w_1&=&\displaystyle\frac{2w_3}{w_1^2 + w_2^2+4},\\ [0.35cm]
\dot w_2&=& \displaystyle\frac{2w_4}{w_1^2 + w_2^2+4},\\ [0.35cm]
\dot w_3 &=& \displaystyle{\frac{2a_2w_1w_2 +a_1(w_1^2 -w_2^2 - 4) +  2a_3w_1 + 2w_1(w_3^2 + w_4^2)}{(w_1^2 + w_2^2 + 4)^2}},\\[0.35cm]
\dot w_4 &=& \displaystyle{\frac{2a_1w_1w_2 - a_2(w_1^2 -w_2^2- 4) + 2a_3w_2 + 2w_2(w_3^2 + w_4^2)}{(w_1^2 + w_2^2 + 4)^2}}.
\end{array}  \right. \label{HDIIIAeq}
\end{equation}
We apply the reduction method \cite{kepler} by choosing  $w_2$ as a new independent variable $y$ which gives rise to the following three equations:
\begin{equation} \left\{
\begin{array}{rcl}
 \displaystyle\der{w_1}{y}&=&\displaystyle\frac{w_3}{w_4},\\ [0.35cm]
\displaystyle\der{w_3}{y} &=&\displaystyle{\frac{2a_2w_1y +a_1(w_1^2 -y^2 - 4) +  2a_3w_1 + 2w_1(w_3^2 + w_4^2)}{2w_4(w_1^2 + y^2 + 4)}},\\ [0.35cm]
\displaystyle\der{w_4}{y} &=&\displaystyle{\frac{2a_1w_1y - a_2(w_1^2 -y^2- 4) + 2a_3y + 2w_2(w_3^2 + w_4^2)}{2w_4(w_1^2 + y^2 + 4)}}.
\end{array}  \right. \label{HDIIIAeqred}
\end{equation}
From the Hamiltonian ${\cal H}_{DIIIA}$, i.e.
\begin{equation}
{\cal H}_{DIIIA}=\displaystyle\frac{w_3^2+w_4^2+a_1w_1+a_2y+a_3}{4+w_1^2+y^2}=h_0,
\end{equation}
we can derive:
\begin{equation}
w_3=\pm\sqrt{h_0(w_1^2+ y^2) + 4h_0 - a_1w_1 - a_2y - a3 - w_4^2},
\end{equation}
with $h_0$ an arbitrary constant. Consequently, the third equation in \eqref{HDIIIAeqred} becomes:
\begin{equation}
\displaystyle\der{w_4}{y}=\displaystyle\frac{ 2h_0y- a_2}{2w_4},
\end{equation}
 that can be easily integrated, i.e.:
 \begin{equation}
w_4=\pm\displaystyle\sqrt{a_2(w_0 - y)+ h_0y^2},
\end{equation}
 with $w_0$ an arbitrary constant. Finally,  we are left with the first equation in \eqref{HDIIIAeqred}, i.e.
 \begin{equation}
 \displaystyle\der{w_1}{y}=\displaystyle\frac{\sqrt{h_0(w_1^2+4)- a_1w_1 - a_2w_0 - a_3 }}{\sqrt{a_2(w_0 - y)+ h_0y^2}}
 \end{equation}
 which can be solved by quadratures. However, if we solve it with respect to $a_3$, and derive once by $y$, then the following linear second-order equation is obtained:
 \begin{equation}
 2\left((w_0-y)a_2+h_0y^2\right)\displaystyle\der[2]{w_1}{y}+(a_2-2h_0y)\displaystyle\der{w_1}{y}+2h_0w_1-w_0-a_1=0.
 \end{equation}

\subsection{Case (B)}
 The
Hamiltonian
\begin{equation}
{\cal H}_{DIIIB}=\displaystyle\frac{w_3^2+w_4^2+\frac{b_1}{w_1^2}+\frac{b_2}{w_2^2}+b_3}{4+w_1^2+w_2^2}
\end{equation}
yields the Hamiltonian equations
\begin{equation} \left\{
\begin{array}{rcl}
\dot w_1&=&\displaystyle\frac{2w_3}{w_1^2 + w_2^2+4},\\ [0.37cm]
\dot w_2&=& \displaystyle\frac{2w_4}{w_1^2 + w_2^2 +4},\\ [0.37cm]
\dot w_3 &=& 2\displaystyle{\frac{\left(b_2 + b_3w_2^2 + (w_3^2 + w_4^2)w_2^2\right)w_1^4 + \left(w_2^2 + 4 + 2w_1^2\right)b_1w_2^2}{(w_1^2 + w_2^2 + 4)^2w_1^3w_2^2}},\\[0.37cm]
\dot w_4 &=& 2\displaystyle{\frac{\left(b_1 + b_3w_1^2 + (w_3^2 + w_4^2)w_1^2\right)w_2^4 + \left(2(w_2^2 + 2) + w_1^2\right)b_2w_1^2}{(w_1^2 + w_2^2 + 4)^2w_1^2w_2^3}}.
\end{array}  \right. \label{HDIIIBeq}
\end{equation}
Before applying the reduction method  \cite{kepler}, we introduce the following transformations of dependent variables, in order to render the next calculations more amenable to a computer algebraic software such REDUCE, i.e.:
\begin{equation}w_1=\sqrt{r_1},\; w_2=\pm\sqrt{r_2},\end{equation}
and then choose  $r_2$ as a new independent variable $y$ which gives rise to the following three equations:
\begin{equation} \left\{
\begin{array}{rcl}
 \displaystyle\der{r_1}{y}&=&\displaystyle{\frac{\sqrt{r_1}w_3}{\sqrt{y}w_4}},\\ [0.37cm]
\displaystyle\der{w_3}{y} &=&\displaystyle{\frac{\left(b_2 + b_3y + (w_3^2 + w_4^2)y\right)r_1^2 + \left(y + 4 + 2r_1\right)b_1y}{2yr_1\sqrt{yr_1}(r_1 + y + 4)w_4}},\\ [0.37cm]
\displaystyle\der{w_4}{y} &=&\displaystyle{\frac{\left(b_1 + b_3r_1 + (w_3^2 + w_4^2)r_1\right)y^2 + \left(2(y + 2) + r_1\right)b_2r_1}{2y^2r_1(r_1 + y + 4)w_4}}.
\end{array}  \right. \label{HDIIIBeqred}
\end{equation}
From the Hamiltonian ${\cal H}_{DIIIB}$, i.e.
\begin{equation}
{\cal H}_{DIIIB}=\displaystyle\frac{w_3^2+w_4^2+\frac{b_1}{r_1}+\frac{b_2}{y}+b_3}{4+r_1+y}=h_0,
\end{equation}
we can derive:
\begin{equation}
w_3=\pm\displaystyle{\sqrt{\frac{\left(h_0(y + 4 + r_1) - b_3 - w_4^2\right)yr_1 - b_2r_1 - b_1y}{yr_1}}},
\end{equation}
with $h_0$ an arbitrary constant. Consequently, the third equation in \eqref{HDIIIBeqred} becomes:
\begin{equation}
\displaystyle\der{w_4}{y}=\displaystyle\frac{b_2+h_0y^2}{2w_4y^2},
\end{equation}
 that can be easily integrated, i.e.:
 \begin{equation}
w_4=\pm\displaystyle{\sqrt{\frac{h_0y(y-w_0)-b_2(1+w_0y)}{y}}},
\end{equation}
 with $w_0$ an arbitrary constant. Finally,  we are left with the first equation in \eqref{HDIIIBeqred}, i.e.
 \begin{equation}
 \displaystyle\der{r_1}{y}=\sqrt{\displaystyle{\frac{b_1 +(b_3- b_2w_0)r_1 - h_0r_1(r_1+w_0 + 4)}{b_2(1+w_0y)+ h_0y(w_0- y)}}},
  \end{equation}
 which can be solved by quadratures. However, if we solve it with respect to $b_1$, and derive once by $y$, then the following linear second-order equation is obtained:
 \begin{equation}
\displaystyle\der[2]{r_1}{y}=-\displaystyle{\frac{ (b_2w_0 + h_0w_0 - 2h_0y)\displaystyle\der{r_1}{y}+ 2h_0r_1+ b_2w_0 - b_3 + h_0w_0 + 4h_0}{ 2\left(b_2(1+w_0y) + h_0y(w_0- y)\right)}}.
 \end{equation}

\subsection{Case (C)}
 The
Hamiltonian
\begin{equation}
{\cal H}_{DIIIC}=\displaystyle\frac{w_1^2w_3^2-w_2^2w_4^2+c_1(w_1+w_2)+c_2\frac{w_1+w_2}{w_1w_2}+c_3\frac{w_1^2-w_2^2}{w_1^2w_2^2}}{(w_1+w_2)(2+w_1-w_2)}
\end{equation}
yields the Hamiltonian equations
\begin{equation} \left\{
\begin{array}{rcl}
\dot w_1&=&\displaystyle\frac{2w_1^2w_3}{(w_1+w_2)(2+w_1-w_2)},\\ [0.37cm]
\dot w_2&=& -\displaystyle\frac{2w_2^2w_4}{(w_1+w_2)(2+w_1-w_2)},\\ [0.37cm]
\dot w_3 &=& 2\displaystyle\frac{(w_3^2 - w_4^2)w_1w_2^2 - 2w_3^2w_1w_2 - w_1^2w_3^2 - w_2^2w_4^2}{(w_1+w_2)^2(2+w_1-w_2)^2}+\frac{c_1}{(2+w_1-w_2)^2}\\[0.37cm]
&&+\displaystyle c_2\frac{2w_1 - w_2 + 2}{(2+w_1-w_2)^2w_1^2w_2}+2c_3\frac{w_1^2 - 2w_1w_2 + w_1 + w_2^2 - 2w_2}{(2+w_1-w_2)^2w_1^3w_2^2},\\[0.37cm]
\dot w_4 &=& -2\displaystyle\frac{(w_3^2 - w_4^2)w_1^2w_2 - w_3^2w_1^2 - (2w_1 + w_2)w_2w_4^2}{(w_1+w_2)^2(2+w_1-w_2)^2} - \frac{c_1}{(2+w_1-w_2)^2}\\[0.37cm]
&& +\displaystyle c_2\frac{w_1 - 2w_2 + 2}{(2+w_1-w_2)^2w_1w_2^2}+2c_3\frac{w_1^2 - 2w_1w_2 + 2w_1 + w_2^2 - w_2}{(2+w_1-w_2)^2w_1^2w_2^3}.
\end{array}  \right. \label{HDIIICeq}
\end{equation}
We apply the reduction method \cite{kepler} by choosing  $w_2$ as a new independent variable $y$ which gives rise to the following three equations:
\begin{equation} \left\{
\begin{array}{rcl}
 \displaystyle\der{w_1}{y}&=&-\displaystyle\frac{w_1^2w_3}{y^2w_4},\\ [0.35cm]
\displaystyle\der{w_3}{y} &=&-\displaystyle\frac{(w_3^2 - w_4^2)w_1y^2 - 2w_3^2w_1y - w_1^2w_3^2 - y^2w_4^2}{(w_1+y)(2+w_1-y)y^2w_4}-c_1\frac{w_1+y}{2(2+w_1-y)y^2w_4}\\[0.37cm]
&&-\displaystyle c_2\frac{(2w_1 - y + 2)(w_1+y)}{2(2+w_1-y)w_1^2y^3w_4}-c_3\frac{(w_1^2 - 2w_1y + w_1 + y^2 - 2y)(w_1+y)}{(2+w_1-y)w_1^3y^4w_4},\\[0.37cm]
\displaystyle\der{w_4}{y} &=&\displaystyle\frac{w_1^2y(w_3^2-w_4^2)-w_1^2w_3^2-2w_1w_4^2y-w_4^2y^2}{(w_1 + y)(2+w_1-y)w_4y^2}+c_1\frac{w_1 + y}{2(2+w_1-y)w_4y^2}\\[0.37cm]
&&\displaystyle -c_2 \frac{(w_1 + y)(w_1 - 2y + 2)}{2(2+w_1-y)w_1w_4y^4} -c_3\frac{ (w_1^2 - 2w_1y + 2w_1 + y^2 - y)(w_1 + y)}{(2+w_1-y)w_1^2w_4y^5}.
\end{array}  \right. \label{HDIIICeqred}
\end{equation}
From the Hamiltonian ${\cal H}_{DIIIC}$, i.e.
\begin{equation}
{\cal H}_{DIIIC}=\displaystyle\frac{w_1^2w_3^2-y^2w_4^2+c_1(w_1+y)+c_2\frac{w_1+y}{w_1y}+c_3\frac{w_1^2-y^2}{w_1^2y^2}}{(w_1+y)(2+w_1-y)}=h_0,
\end{equation}
we can derive:
\begin{equation}
w_3=\pm\displaystyle{\frac{\sqrt{w_1^2w_4^2y^4-w_1y(w_1+y)(c_1w_1y+c_2)+c_3(y^2-w_1^2)+h_0(w_1+y)(2+w_1-y)w_1^2y^2}}{w_1^2y}},
\end{equation}
with $h_0$ an arbitrary constant. Consequently, the third equation in \eqref{HDIIICeqred} becomes:
\begin{equation}
\displaystyle\der{w_4}{y}=\displaystyle\frac{2w_4^2y^4-c_1y^3+c_2y+2c_3+2y^3h_0(1-y)}{2w_4y^5},
\end{equation}
 that can be easily integrated, i.e.:
 \begin{equation}
w_4=\pm\displaystyle\frac{\sqrt{w_0y^2+c_1y^3+c_2y+c_3+h_0y^3(y-2)}}{y^2},
\end{equation}
 with $w_0$ an arbitrary constant.  Finally,  we are left with the first equation in \eqref{HDIIICeqred}, i.e.
 \begin{equation}
 \displaystyle\der{w_1}{y}=-\displaystyle\frac{\sqrt{w_0w_1^2-c_1w_1^3-c_2w_1+c_3+h_0w_1^3(w_1+2)}}{\sqrt{w_0y^2+c_1y^3+c_2y+c_3+h_0y^3(y-2)}},\label{HIIIC1ode}
 \end{equation}
   which could be solved by quadratures. If we introduce new parameters in order to simplify this equation, i.e.:
 \begin{equation}
 c_1=C_1+2h_0,\quad  \quad  c_2=C_2C_1, \quad c_3=C_3C_1, \quad h_0=H_0C_1,\quad w_0=W_0C_1,
 \end{equation}
 and the new dependent variable $u=-w_1$, then equation \eqref{HIIIC1ode} becomes:
\begin{equation}
 u'(y)\equiv \displaystyle\der{u}{y}=\displaystyle\frac{\sqrt{C_2u + C_3 + H_0u^4 + W_0u^2 + u^3}}{\sqrt{C_2y + C_3 + H_0y^4 + W_0y^2 + y^3}}.\label{HIIIC1odeu}
 \end{equation}
If we solve this first-order equation with respect to $C_3$, and derive once by $y$, then a second-order equation is obtained. If we solve this second-order equation with respect to $W_0$, and derive once by $y$,  then a third-order equation is obtained. Finally, if we solve this third-order equation with respect to $H_0$, and derive once by $y$, then the following fourth-order equation is obtained:
\begin{equation}
    u^{(iv)}=-\frac{%
        \alpha_{1} u''^3
        +\alpha_{2} u''^2
        -\alpha_{3} u'' u'''
        -\alpha_{4} u''
        +\alpha_{5} u'''^2
        -6\alpha_{7} u'''
        +\alpha_{8}
    }{3 (C_2 - uy) (u u'' - 2 u'^2 - 2 u' - u'' y) (u^2 - y^2)},
    \label{DIIICo4}
\end{equation}
with
\begin{eqnarray}
          \alpha_{1} &=&9 (u-y) [C_2 (3  u+5 y)-2 u^2 y-5 u y^2-y^3],
    \nonumber
        \\
        \alpha_{2} & =& C_{2}(36 u+54 y +36 u u'-54 u'^2 y)
        -36 u^2 u'^2 y+72 u u'^2 y^2+18 u'^2 y^3\nonumber\\
       &&+18 u^3 u'-72 u^2 u' y-18 u u' y^2+36 u' y^3
        -18 u^2 y-72 u y^2,
         \nonumber
        \\
        \alpha_{3} &=&3 (u-y) \left[
                           C_{2} (13 u u'+15 u' y+5 u+7 y)-12 u^2 u' y
                            -15 u u' y^2-u' y^3
            +u^3-5 u^2 y-8 u y^2
                  \right],
         \nonumber
        \\
        \alpha_{4} &=&18 u' (u'+1) \left[
                           C_2 (u'^2-1)-4 u u'^2 y+u'^2 y^2
                         +3 u^2 u'-3 u' y^2-u^2+4 u y
                \right],
       \nonumber
        \\
        \alpha_{5} &=& 5 (u+y) (u-y)^2 (C_2-u y),
        \nonumber
        \\
        \alpha_{7}&=& u' (u'+1) \left[
                           C_2(3  u u'+5 u' y-5 u-3 y)-2 u^2 u' y
                                -5 u u' y^2-u' y^3+u^3+5 u^2 y+2 u y^2
              \right], \nonumber
        \\
        \alpha_{8}&=&36 u'^2 (u'-1) (u'+1)^2 (u-u' y).
   \nonumber
\end{eqnarray}
It admits a fourth-dimensional Lie symmetry algebra $2A_2$ generated by the following operators:
\begin{eqnarray}
  \Gamma_1&=& \frac{3}{u-y}\left(-(C_2^2+u y^3)\partial_y+(C_2^2+u^3y)\partial_u\right), \\
  \Gamma_2 &=& \frac{u+y}{u-y}\left(-(C_2+y^2)\partial_y+(u^2+C_2)\partial_u\right),\\
  \Gamma_3 &=& \frac{1}{u-y}\left((u y-2C_2-y^2)\partial_y+(2C_2+u^2-u y)\partial_u\right),\\
  \Gamma_4 &=& -\frac{1}{3(u-y)}\left((C_2-2u y-y^2)\partial_y+(u^2-C_2+2u y)\partial_u\right).
  \end{eqnarray}
  In order to follow the classification of the fourth-dimensional Lie symmetry algebra in \cite{Patera}, and the fourth-order equations admitting them as derived in \cite{cercicnuc}, we choose another representation of the operators that generate $2A_2$, i.e.:
  \begin{equation}
   X_1= \Gamma_1-3C_2\Gamma_3, \quad X_2=\Gamma_2, \quad X_3 = \frac{1}{3}\Gamma_3+\Gamma_4,\quad  X_4 = \frac{1}{3}\Gamma_3-2\Gamma_4.
   \end{equation}
   We thank Nicola Ciccoli for his invaluable help on this issue.
A two-dimensional Abelian intransitive subalgebra of the Lie symmetry algebra $2A_2$ is that generated by $X_1$ and $X_2$, and the corresponding canonical transformations \cite{Lie12} are:
 $$\tilde y= \frac{u+y}{3(u y-C_2)}, \quad \tilde u= -\frac{1}{u y-3C_2}. $$
Then equation \eqref{DIIICo4} turns into the following fourth-order equation:
 \begin{equation}
\displaystyle \frac{{\rm d}^4\tilde u}{{\rm d}\tilde y^4}=\left(3\frac{{\rm d}^2\tilde u}{{\rm d}\tilde y^2}+5\tilde y \frac{{\rm d}^4\tilde u}{{\rm d}\tilde y^4}\right)\displaystyle\frac{\displaystyle\frac{{\rm d}^3\tilde u}{{\rm d}\tilde y^3}}{3\tilde y \displaystyle\frac{{\rm d}^2\tilde u}{{\rm d}\tilde y^2}}.\label{DIIICo4tr}
 \end{equation}
 If we make the substitution
 $$\displaystyle\frac{{\rm d}^2\tilde u}{{\rm d}\tilde y^2}=R(\tilde y),$$
 then equation \eqref{DIIICo4tr} becomes the following second-order equation:
  \begin{equation}
  \displaystyle \frac{{\rm d}^2 R}{{\rm d}\tilde y^2}=\left(3R+5\tilde y \frac{{\rm d}R}{{\rm d}\tilde y}\right)\displaystyle\frac{\displaystyle\frac{{\rm d}R}{{\rm d}\tilde y}}{3\tilde y R},\label{DIIICo2R}
  \end{equation}
  which admits an eight-dimensional Lie symmetry algebra $sl(3,\R)$, and therefore it is linearizable \cite{Lie12}. Indeed, the transformation $Y= R^{2/3},\;U=\displaystyle \frac{\tilde y^2}{2}R^{2/3}$ yields $$ \displaystyle \frac{{\rm d}^2 U}{{\rm d}Y^2}=0 \Rightarrow U=A_1Y +A_2,$$
  with $A_1,A_2$ arbitrary constants.
  Consequently, the general solution of equation \eqref{DIIICo2R} is
    \begin{equation}
 R=\displaystyle\frac{2A_2\sqrt{2A_2}}{(\tilde y^2-2A_1)\sqrt{\tilde y^2-2A_1}},
  \end{equation}
  that integrated twice yields the general solution of equation \eqref{DIIICo4tr}, i.e.:
  \begin{equation}
\tilde u=\frac{A_2}{A_1}\sqrt{2A_2(\tilde y^2-2A_1)}+A_3\tilde y+A_4,
  \end{equation}
  with $A_3,A_4$ arbitrary constants. Finally, the general solution of equation \eqref{DIIICo4} is:
  \begin{equation}
u=\displaystyle\frac{ \beta_2 y^2+\beta_1y+\beta_0+\sqrt{\gamma_4y^4+ \gamma_3 y^3+\gamma_2y^2+\gamma_1 y+\gamma_0}}{9 A_1 (A_4^2 A_1+4 A_2^3)y^2+6 A_1^2 A_3 A_4 y+A_1^2 A_3^2-2 A_2^3},
  \end{equation}
  with
  \begin{eqnarray}
  \beta_2 &=& -3 A_1^2 A_3 A_4,\nonumber\\
           \beta_1 &=& 9 A_1^2 A_4^2 C_2-3 A_1^2 A_4 +2 A_2^3-A_1^2 A_3^2+36 A_2^3 A_1 C_2,\nonumber\\
           \beta_0 &=& A_1^2 A_3 (3 C_2 A_4-1),\nonumber\\
  \gamma_4 &=& -18 A_1^2 A_3^2+9 A_4^2 A_1+36 A_2^3,\nonumber\\
    \gamma_3 &=& -36 A_1^2 A_3,\nonumber\\
      \gamma_2 &=& 18 A_4^2 A_1 C_2-6 A_4 A_1+72 A_2^3 C_2-36 A_1^2 A_3^2 C_2-18 A_1^2,\nonumber\\
        \gamma_1 &=& -36 A_1^2 C_2 A_3,\nonumber\\
          \gamma_0 &=& -18 A_1^2 A_3^2 C_2^2+9 A_1 A_4^2 C_2^2-6 A_1 C_2 A_4+A_1+36 A_2^3 C_2^2.\nonumber
  \end{eqnarray}
We would like to remark that if we  solve the fourth-order equation \eqref{DIIICo4} with respect to $C_2$, and derive once by $y$, then a fifth-order equation is obtained, which admits an eight-dimensional Lie symmetry algebra, and can be transformed into a third-order linearizable equation since it admits a seven-dimensional Lie symmetry algebra, quite similarly to the fourth-order equation that we discuss in details here.

\subsection{Case (D)}
 The
Hamiltonian
\begin{equation}
{\cal H}_{DIIID}=\displaystyle\frac{w_1^2w_3^2-w_2^2w_4^2+d_1w_1+d_2w_2+d_3(w_1^2+w_2^2)}{(w_1+w_2)(2+w_1-w_2)}
\end{equation}
yields the Hamiltonian equations
\begin{equation} \left\{
\begin{array}{rcl}
\dot w_1&=&\displaystyle\frac{2w_1^2w_3}{(w_1+w_2)(2+w_1-w_2)},\\ [0.37cm]
\dot w_2&=& -\displaystyle\frac{2w_2^2w_4}{(w_1+w_2)(2+w_1-w_2)},\\ [0.37cm]
\dot w_3 &=& \displaystyle{\frac{1}{(w_1+w_2)^2(2+w_1-w_2)^2}}\Big(2\left(w_1 w_2^2(w_3^2-w_4^2)-2w_1 w_2 w_3^2 -w_2^2w_4^2-w_1^2 w_3^2\right)\\&&+d_1(w_1^2+w_2^2-2w_2)+2d_2w_2(w_1+1)+2d_3(w_2^2-w_1^2+2w_1w_2^2-2w_1w_2)\Big),\\[0.37cm]
\dot w_4 &=& \displaystyle{\frac{1}{(w_1+w_2)^2(2+w_1-w_2)^2}}\Big(2\left(w_1^2 w_2(w_4^2-w_3^2)+2w_1 w_2 w_4^2 +w_2^2w_4^2+w_1^2 w_3^2\right)\\&&+2d_1w_1(1-w_2)-d_2(w_1^2+w_2^2+2w_1)-2d_3(w_2^2-w_1^2+2w_1^2w_2+2w_1w_2)\Big).
\end{array}  \right. \label{HDIIIDeq}
\end{equation}
We apply the reduction method \cite{kepler} by choosing  $w_2$ as a new independent variable $y$ which gives rise to the following three equations:
\begin{equation} \left\{
\begin{array}{rcl}
 \displaystyle\der{w_1}{y}&=&-\displaystyle\frac{w_1^2w_3}{y^2w_4},\\ [0.35cm]
\displaystyle\der{w_3}{y} &=&-\displaystyle{\frac{1}{2w_4y^2(w_1+y)(2+w_1-y)}}\Big(2\left(w_1 y^2(w_3^2-w_4^2)-2w_1 y w_3^2 -y^2w_4^2-w_1^2 w_3^2\right)\\&&+d_1(w_1^2+y^2-2y)+2d_2y(w_1+1)+2d_3(y^2-w_1^2+2w_1y^2-2w_1y)\Big),\\[0.37cm]
\displaystyle\der{w_4}{y} &=&-\displaystyle{\frac{1}{2w_4y^2(w_1+y)(2+w_1-y)}}\Big(2\left(w_1^2 y(w_4^2-w_3^2)+2w_1 y w_4^2 +y^2w_4^2+w_1^2 w_3^2\right)\\&&+2d_1w_1(1-y)-d_2(w_1^2+y^2+2w_1)-2d_3(y^2-w_1^2+2w_1^2y+2w_1y)\Big).
\end{array}  \right. \label{HDIIIDeqred}
\end{equation}
From the Hamiltonian ${\cal H}_{DIIID}$, i.e.
\begin{equation}
{\cal H}_{DIIID}=\displaystyle\frac{w_1^2w_3^2-y^2w_4^2+d_1w_1+d_2y+d_3(w_1^2+y^2)}{(w_1+y)(2+w_1-y)}=h_0,
\end{equation}
we can derive:
\begin{equation}
w_3=\pm\displaystyle{\frac{\sqrt{(h_0-d_3)w_1^2+(2h_0-d_1)w_1+(2h_0-d_2)y-(d_3+h_0)y^2+w_4^2y^2}}{w_1}},
\end{equation}
with $h_0$ an arbitrary constant. Consequently, the third equation in \eqref{HDIIIDeqred} becomes:
\begin{equation}
\displaystyle\der{w_4}{y}=\displaystyle\frac{d_2-2h_0+ 2(d_3+h_0)y- 2w_4^2y}{2w_4y^2},
\end{equation}
 that can be easily integrated, i.e.:
 \begin{equation}
w_4=\pm\displaystyle{\frac{\sqrt{(d_2-2h_0)y+ (d_3+ h_0)y^2+w_0}}{y}},
\end{equation}
 with $w_0$ an arbitrary constant. Let us introduce new parameters that simplify the formula for $w_3$ and $w_4$, i.e.:
 \begin{equation}
 D_1=2h_0-d_1,\quad  \quad  D_2=d_2 - 2h_0, \quad \quad D_3=d_3 + h_0,
 \end{equation}
 and consequently:
 \begin{equation}
 w_3=\pm\displaystyle{\frac{\sqrt{(2h_0-D_3)w_1^2+D_1w_1+w_0}}{w_1}},\quad w_4=\pm\displaystyle{\frac{\sqrt{D_2y+ D_3y^2+w_0}}{y}}.
 \end{equation}
  Finally,  we are left with the first equation in \eqref{HDIIIDeqred}, i.e.
 \begin{equation}
 \displaystyle\der{w_1}{y}=-\displaystyle\frac{w_1\sqrt{(2h_0-D_3)w_1^2+D_1w_1+w_0}}{y\sqrt{D_3y^2+D_2y+w_0}},
 \end{equation}
  which can be solved by quadratures. However, if we solve it with respect to $D_1$, and derive once by $y$, then the following  second-order equation is obtained:
 \begin{eqnarray}
 2y^2w_1(D_3y^2+D_2y+w_0)\displaystyle\der[2]{w_1}{y}-3y^2(D_3y^2+D_2y+w_0) \left(\displaystyle\der{w_1}{y}\right)^2\nonumber\\+(4D_3y^2+3D_2y+2w_0)yw_1\displaystyle\der{w_1}{y}+(D_3-2h_0)w_1^4+w_0w_1^2=0,\label{HIIIDeqo2}
 \end{eqnarray}
that admits a three-dimensional symmetry algebra $\Sl(2,\R)$, unless $D_3=2h_0$ in which case it admits an eight-dimensional Lie symmetry algebra $\Sl(3,\R)$ and thus it is linearizable. We now use the general method described in \cite{Leach2003} and that may be
applied to any second-order ordinary differential equation that admits a Lie
symmetry algebra $\Sl(2,\R)$. If we solve equation \eqref{HIIIDeqo2} with respect to
$h_0$ and derive once with respect to $y$, then we obtain the following third-order equation:
\begin{eqnarray}
\frac{2yw_1^2}{3}(D_3y^2+D_2y+w_0)\displaystyle\der[3]{w_1}{y}= -4y(D_3y^2+D_2y+w_0) \left(\displaystyle\der{w_1}{y}\right)^3 \nonumber \\ +2w_1(4D_3y^2+3D_2y+2w_0) \left(\displaystyle\der{w_1}{y}\right)^2-2(2D_3y+D_2)w_1^2  \displaystyle\der{w_1}{y}  \nonumber \\ +\left(4yw_1(D_3y^2+D_2y+w_0)\displaystyle\der{w_1}{y}-(4D_3y^2+3D_2y+2w_0)w_1^2  \right)\displaystyle\der[2]{w_1}{y}, \label{HIIIDeqo3}
\end{eqnarray}
which admits a seven-dimensional Lie symmetry algebra, and therefore is linearizable. Indeed, the new dependent and independent variables, i.e. \begin{equation}\tilde w_1=-\frac{1}{w_1},\quad\quad \tilde y= \frac{2D_3y+D_2}{y},\end{equation} transform equation \eqref{HIIIDeqo3} into the linear equation
\begin{equation}\displaystyle\der[3]{\tilde w_1}{\tilde y}=\displaystyle\frac{3(4D_3w_0-D_2^2-2w_0\tilde y)}{2\left(w_0\tilde y^2+(4D_3w_0-D_2^2)(D_3-\tilde y)\right)}\displaystyle\der[2]{\tilde w_1}{\tilde y}.\end{equation}
\subsection{Case (E)}
 The
Hamiltonian
\begin{equation}
{\cal H}_{DIIIE}=\frac{w_3^2+w_4^2+c}{4+w_1^2+w_2^2}
\end{equation}
 is a subcase of Hamiltonian ${\cal H}_{DIIIA}$, with $a_1=a_2=0$ and $a_3=c$. Consequently, its corresponding Hamiltonian equations, i.e.:
 \begin{equation} \left\{
\begin{array}{rcl}
\dot w_1&=&\displaystyle\frac{2w_3}{w_1^2 + w_2^2+4},\\ [0.35cm]
\dot w_2&=& \displaystyle\frac{2w_4}{w_1^2 + w_2^2+4},\\ [0.35cm]
\dot w_3 &=& \displaystyle{\frac{2w_1(c+w_3^2 + w_4^2)}{(w_1^2 + w_2^2 + 4)^2}},\\[0.35cm]
\dot w_4 &=& \displaystyle{\frac{2w_2(c+w_3^2 + w_4^2)}{(w_1^2 + w_2^2 + 4)^2}}.
\end{array}  \right. \label{HDIIIEeq}
\end{equation}
can be reduced to the following linear equation in $w_1=w_1(w_2)$:
 \begin{equation}
 (2w_0-w_2^2)\displaystyle\der[2]{w_1}{w_2}-w_2\displaystyle\der{w_1}{w_2}+w_1=0,
  \end{equation}
  with
\begin{equation}
w_3=\pm\sqrt{h_0(w_1^2+ w_2^2) + 4h_0 - c - w_4^2},
\end{equation}
 and
 \begin{equation}
w_4=\pm\displaystyle \sqrt{h_0(w_2^2-2w_0)}.
\end{equation}

\section{Darboux IV}
Four superintegrable systems were determined in
\cite{KalKreMillWint}, where the problem of superintegrability for the Hamiltonian
\begin{equation}
H_{DIV}=-\sin^2(2u)\frac{p_u^2+p_v^2}{2\cos(2u)+a}
\end{equation}
was addressed.
\subsection{Case (A)}
 The
Hamiltonian
\begin{equation}
{\cal H}_{DIVA}=-4w_1^2w_2^2\displaystyle\frac{w_3^2+w_4^2+a_1+a_2\left(\frac{1}{w_1^2}+\frac{1}{w_2^2}\right)+a_3(w_1^2+w_2^2)}{(a+2)w_1^2+(a-2)w_2^2}
\end{equation}
yields the Hamiltonian equations
\begin{equation} \left\{
\begin{array}{rcl}
\dot w_1&=&-\displaystyle\frac{8w_1^2w_2^2w_3}{a(w_1^2 + w_2^2) + 2(w_1^2-w_2^2)},\\ [0.37cm]
\dot w_2&=& -\displaystyle\frac{8w_1^2w_2^2w_4}{a(w_1^2 + w_2^2) + 2(w_1^2-w_2^2)},\\ [0.37cm]
\dot w_3 &=& \displaystyle{\frac{8w_1w_2^2}{\left(a(w_1^2 + w_2^2) + 2(w_1^2-w_2^2)\right)^2}}\Big(aa_3(w_1^2+w_2^2)^2 +aw_2^2(w_3^2 +w_4^2) +a_1w_2^2(a - 2) - 4a_2\\&&+ 2a_3(w_1^4-w_2^4-2w_1^2w_2^2)  - 2w_2^2(w_3^2+w_4^2)\Big),\\[0.37cm]
\dot w_4 &=& \displaystyle{\frac{8w_1^2w_2}{\left(a(w_1^2 + w_2^2) + 2(w_1^2-w_2^2)\right)^2}}\Big(aa_3(w_1^2+w_2^2)^2+aw_1^2(w_3^2 +w_4^2)+a_1w_1^2(a+2) + 4a_2\\ &&+
2a_3(w_1^4-w_2^4+2w_1^2w_2^2)+ 2w_1^2(w_3^2+w_4^2)\Big).
\end{array}  \right. \label{HDIVAeq}
\end{equation}
In order to simplify the calculations, we make the following substitutions of the four dependent variables:
\begin{equation}
w_1=\sqrt{r_1},\quad w_2=\sqrt{r_2},\quad w_3=\sqrt{r_3},\quad w_4=\sqrt{r_4}, \label{allroots}
\end{equation}
and consequently system \eqref{HDIVAeq} is transformed into the following system:
\begin{equation} \left\{
\begin{array}{rcl}
\dot r_1&=&-\displaystyle\frac{16r_1\sqrt{r_1r_3}r_2}{a(r_1+r_2) + 2(r_1-r_2)},\\ [0.37cm]
\dot r_2&=& -\displaystyle\frac{16r_2\sqrt{r_2r_4}r_1}{a(r_1+r_2) + 2(r_1-r_2)},\\ [0.37cm]
\dot r_3 &=& \displaystyle{\frac{16\sqrt{r_1r_3}r_2}{\left(a(r_1+r_2) + 2(r_1-r_2)\right)^2}}\Big(aa_3(r_1+r_2)^2 +ar_2(r_3 +r_4) +a_1r_2(a - 2) - 4a_2\\&&
+ 2a_3(r_1^2-r_2^2-2r_1r_2)  - 2r_2(r_3+r_4)\Big),\\[0.37cm]
\dot r_4 &=& \displaystyle{\frac{16\sqrt{r_2r_4}r_1}{\left(a(r_1+r_2) + 2(r_1-r_2)\right)^2}}\Big(aa_3(r_1+r_2)^2++ar_1(r_3 +r_4)+a_1r_1(a+2) + 4a_2\\ &&
+ 2a_3(r_1^2-r_2^2+2r_1r_2)+ 2r_1(r_3+r_4)\Big).
\end{array}  \right. \label{HDIVAeq2}
\end{equation}
We apply the reduction method \cite{kepler} by choosing  $r_2$ as a new independent variable $y$ which gives rise to the following three equations:
\begin{equation} \left\{
\begin{array}{rcl}
 \displaystyle\der{r_1}{y}&=&\sqrt{\displaystyle\frac{r_1r_3}{yr_4}},\\ [0.35cm]
\displaystyle\der{r_3}{y} &=&-\displaystyle{\frac{\sqrt{r_1r_3}}{r_1\sqrt{yr_4}\left(a(r_1+y) + 2(r_1-y)\right)}}\Big(aa_3(r_1+y)^2 +ay(r_3 +r_4) +a_1y(a - 2) - 4a_2\\&&
+ 2a_3(r_1^2-y^2-2r_1y)  - 2y(r_3+r_4)\Big),\\[0.37cm]
\displaystyle\der{r_4}{y} &=&\displaystyle{\frac{1}{y\left(a(r_1+y) + 2(r_1-y)\right)}}\Big(aa_3(r_1+y)^2+ar_1(r_3 +r_4)+a_1r_1(a+2) + 4a_2\\ &&
+ 2a_3(r_1^2-y^2+2r_1y)+ 2r_1(r_3+r_4)\Big).
\end{array}  \right. \label{HDIVAeqred}
\end{equation}
From the Hamiltonian ${\cal H}_{DIVA}$, i.e.:
\begin{equation}
{\cal H}_{DIVA}=-4r_1y\displaystyle\frac{r_3+r_4+a_1+a_2\left(\frac{1}{r_1}+\frac{1}{y}\right)+a_3(r_1+y)}{(a+2)r_1+(a-2)y}=h_0,
\end{equation}
we can derive:
\begin{equation}
r_3=-\displaystyle{\frac{4r_1r_4y + ((a+ 2)r_1 + (a - 2)y)h_0 + 4a_1r_1y + 4a_2(r_1 + y) + 4a_3r_1 y(r_1 +y)}{4r_1y}},
\end{equation}
with $h_0$ an arbitrary constant. Consequently, the third equation in \eqref{HDIVAeqred} becomes:
\begin{equation}
\displaystyle\der{r_4}{y}=\displaystyle\frac{(a+2)h_0 + 4a_2 - 4a_3y^2}{4y^2},
\end{equation}
  that can be easily integrated, i.e.:
 \begin{equation}
r_4=\displaystyle\frac{4r_0y-(a+2)h_0-4a_2-4a_3y^2}{4y},
\end{equation}
 with $r_0$ an arbitrary constant.  Finally,  we are left with the first equation in \eqref{HDIVAeqred}, i.e.:
 \begin{equation}
 \displaystyle\der{r_1}{y}=\sqrt{\displaystyle\frac{(2-a)h_0 - 4a_2  - 4r_1(a_1 + r_0)- 4a_3r_1^2}{-(a+2)h_0 - 4 a_2 - 4 a_3 y^2  + 4 r_0 y}},\label{HIVA1ode}
 \end{equation}
   which could be easily solved by quadratures. If we introduce a new parameter $b_2=-(a+2)h_0 - 4 a_2$, such that $a_2=-((a+2)h_0+ b_2)/4$, and then solve the first-order equation \eqref{HIVA1ode}  with respect to $a_1$, and derive once by $y$, then a second-order equation is obtained. If we solve this second-order equation with respect to $h_0$, and derive once by $y$,  then the following linear third-order equation is obtained:
   \begin{equation}
   (b_2 + 4r_0y - 4a_3y^2)\displaystyle\der[3]{r_1}{y} - 6(2a_3y - r_0)\displaystyle\der[2]{r_1}{y}=0.
   \end{equation}

\subsection{Case (B)}
The
Hamiltonian
\begin{equation}
{\cal H}_{DIVB}=-\displaystyle\frac{\sin^2(2w_1)\left( w_3^2+w_4^2+\frac{b_2}{\sinh^2(w_2)}+\frac{b_3}{\cosh^2(w_2)} \right)+b_1}{2\cos(2w_1)+a}
\end{equation}
can be written in the following equivalent form with $\sinh$, and $\cosh$ replaced by $\exp$, i.e.:
\begin{equation}
    {\cal H}_{DIVB}=-\displaystyle\frac{\sin^2(2w_1)\left( w_3^2+w_4^2+\frac{4b_2}{\left(e^{w_2}-e^{-w_{2}}\right)}+\frac{4b_3}{\left( e^{w_{2}}+e^{-w_{2}} \right)} \right)+b_1}{2\cos(2w_1)+a}.
\end{equation}
We apply the reduction method \cite{kepler} by choosing  $r_2=e^{w_2}$ as a new independent variable $y$ which gives rise to the following three equations:
\begin{equation}
    \left\{
    \begin{array}{rcl}
        \displaystyle\der{w_1}{y}&=&\displaystyle\frac{w_{3}}{yw_{4}},
        \\
        [0.37cm]
        \displaystyle\der{w_3}{y} &=& \displaystyle\frac{N}{\sin(2w_1)w_4 (y^{4}-1)^2[a+2\cos(2w_1)]},
        \\
        [0.37cm]
        \displaystyle\der{w_4}{y} &=&
        \displaystyle\frac{4y^{2} \left[ (b_2+b_3) y^8+4(b_2-b_3) y^6+6(b_2+b_3) y^4+4(b_2- b_3) y^2+b_2+b_3 \right]
        }{w_{4}\left( y^{4}-1 \right)^{3}},
    \end{array}
    \right.
    \label{HDIVBeqred}
\end{equation}
where
\begin{equation}
    \begin{aligned}
        N &=
        -\left[4(w_3^2+ w_4^2) y^8+16(b_2+ b_3) y^6-8(w_3^2+ w_4^2-4 b_2+4 b_3) y^4
            +16(b_2+ b_3) y^2+4 (w_3^2+ w_4^2)\right] \cos(2 w_1)^2
        \\
        &-2 a \left[(w_3^2+w_4^2) y^8+4(b_2+b_3) y^6-2( w_3^2+ w_4^2-4 b_2-4 b_3) y^4
            +4(b_2+b_3) y^2+w_3^2+w_4^2\right] \cos(2 w_1)
        \\
        &+\left[ -2(w_3^2+ w_4^2) y^8-8( b_2+ b_3) y^6+ 4( w_3^2+ w_4^2-4 b_2+4 b_3) y^4
        -8( b_2+ b_3) y^2-2( w_3^2+ w_4^2)\right] \sin(2 w_1)^2
        \\
        &-2 b_1 y^8+4 b_1 y^4-2 b_1.
    \end{aligned}
    \label{eq:NDIVB}
\end{equation}
We can solve the third equation with respect to $w_4$, i.e.:
\begin{equation}
w_4=\pm \displaystyle\frac{2}{15\sqrt{15}(y^4 - 1)}\sqrt{3375(b_3-b_2)(y^8+1)- 3375(b_3+b_2)(y^4+1)y^2 - 4(625b_2+81b_3)(y^4-1)^2w_0}.
\end{equation}
Then the first equation in \eqref{HDIVBeqred} yields:
\begin{equation}
w_3=yw_4\displaystyle\der{w_1}{y},
\end{equation}
which replaced into the second equation in \eqref{HDIVBeqred} gives rise to a second-order equation in $w_1$ that we solve with respect to $b_1$. The we derive  once with respect to $y$ and the following third-order equation is obtained:
\begin{equation}
    \begin{aligned}
        \der[3]{w_{1}}{y} &=
        -6\cot(2u)\der{w_{1}}{y}\der[2]{w_{1}}{y}
        -\frac{3}{y\left( y^{4}-1 \right)}
        \frac{P_{1}\left( y,u \right)}{Q\left( y,u \right)}\der[2]{w_{1}}{y}
        +4 \left( \der{w_{1}}{y} \right)^3
        \\
        &-\frac{6\cot(2u)}{y\left( y^{4}-1 \right)}
        \frac{P_{1}\left( y,u \right)}{Q\left( y,u \right)}
        \left( \der{w_{1}}{y} \right)^2
        +\frac{3}{y^{2}\left( y^{4}-1 \right)^{2}}
        \frac{P_{2}\left( y \right)}{Q\left( y \right)}\der{w_{1}}{y},
    \end{aligned}
    \label{eq:w1IIIDIVB}
\end{equation}
where:
\begin{align}
    P_{1}&
    \begin{aligned}[t]
        &=[(2500 b_2+324 b_3) w_0+3375 (b_2- b_3)] y^{12}
        -[(7500 b_2+972 b_3) w_0+16875 (b_2- b_3)] y^8
        -20250 (b_2+ b_3) y^6
        \\
        &+[(7500 b_2+972 b_3) w_0-10125 (b_2-b_3)] y^4
        -6750 (b_2+ b_3) y^2
        -(2500 b_2+324 b_3) w_0-3375 (b_2- b_3),
    \end{aligned}
    \label{eq:P1}
    \\
    P_{2} &
    \begin{aligned}[t]
        &=
        [(2500 b_2+324 b_3) w_0+3375 (b_2- b_3)] y^{16}
        -[(10000 b_2+1296 b_3) w_0-20250 (b_2- b_3)] y^{12}
        \\
        &-33750 (b_2+ b_3) y^{10}
        +[(15000 b_2+1944 b_3) w_0-47250 (b_2- b_3)] y^8
        -67500 (b_2+ b_3) y^6
        \\
        &-[(10000 b_2+1296 b_3) w_0-47250 (b_2- b_3)] y^4
        -6750 (b_2+ b_3) y^2
        +(2500 b_2+324 b_3) w_0+3375 (b_2- b_3),
    \end{aligned}
    \label{eq:P2}
    \\
    Q &
    \begin{aligned}[t]
        &=[(2500 b_2+324 b_3) w_0+3375 (b_2- b_3)] y^8
        +3375 (b_2+ b_3) y^6
        -(5000 b_2+648 b_3) w_0 y^4
        \\
        &+3375( b_2+ b_3) y^2
        +(2500 b_2+324 b_3) w_0
        +3375 (b_2- b_3).
    \end{aligned}
    \label{eq:Q11}
\end{align}
Equation \eqref{eq:w1IIIDIVB} is linearizable since it admits a seven-dimensional Lie symmetry algebra. In fact, the two-dimensional abelian intransitive subalgebra generated by the two operators
\begin{equation}
\displaystyle -\frac{\cos(2 u)}{2\sin(2 u)} \partial_{u},\quad \quad \displaystyle\frac{1}{\sin(2 u)}\partial_{u}
\end{equation}
when put into the canonical form $\partial_{\tilde u},\tilde y\partial_{\tilde u}$ yield the new dependent and independent variables, i.e.
\begin{equation}\tilde u=-\frac{1}{2}\cos(2 u),\quad
\tilde y=\displaystyle\frac{-B_2y^4-B_2+4B_3y^2+96W_0y^2}{6y^2},\end{equation}
where we have introduced new constants $B_2,B_3,W_0$ such that:
\begin{equation}
b_2=\displaystyle\frac{B_2 - B_3 - 12W_0}{40500},\quad b_3=\displaystyle\frac{B_2 + B_3 + 12W_0}{40500},
\quad w_0=\displaystyle\frac{3375(B_3 + 18W_0)}{4(353B_2 - 272B_3 - 3264W_0)}.
\end{equation} Then equation \eqref{HIVCeqo3} transforms into the linear equation
\begin{equation}\displaystyle\der[3]{\tilde u}{\tilde y}=\frac{9}{2}\displaystyle\der[2]{\tilde u}{\tilde y}\frac{B_2^2 + 48B_3W_0 + 1152W_0^2 - 72W_0\tilde y}
{B_2^2B_3 + 24(B_2^2+2B_3^2)W_0 + 2304B_3W_0^2+ 27648W_0^3- 3(B_2^2+48B_3W_0+1152W_0^2)\tilde y + 108W_0\tilde y^2}.
\end{equation}

\subsection{Case (C)}
The
original Hamiltonian
\begin{equation}
{\cal H}_{DIVC}=-\displaystyle{\frac{w_3^2+w_4^2+\frac{c_1}{\cos^2(w_1)}+\frac{c_2}{\cosh^2(w_2)}+c_3\left(\frac{1}{\sin^2(w_1)}
-\frac{1}{\sinh^2(w_2)}\right)}{\frac{a+2}{\sinh^2(2w_2)}+\frac{a-2}{\sin^2(2w_1)}}},
\end{equation}
can be written in the following equivalent form with $\sinh$, and $\cosh$ replaced by $\exp$, i.e.:
\begin{equation}
{\cal H}_{DIVC}=-\displaystyle{\frac{w_3^2+w_4^2+\frac{c_1}{\cos^2(w_1)}+\frac{c_2}{\left(\frac{e^{w_2}+e^{-w_2}}{2}\right)^2}+c_3\left(\frac{1}{\sin^2(w_1)}
-\frac{1}{\left(\frac{e^{w_2}-e^{-w_2}}{2}\right)^2}\right)}{\frac{a+2}{\left(\frac{e^{2w_2}-e^{-2w_2}}{2}\right)^2}+\frac{a-2}{\sin^2(2w_1)}}}.
\end{equation}
Before applying the reduction method  \cite{kepler}, we introduce the following transformations of dependent variables, in order to render the next calculations more amenable to a computer algebraic softwares such REDUCE and MAPLE, i.e.:
\begin{equation}w_1=\arccos{r_1},\; w_2=\log{r_2},\; w_3=\sqrt{r_3},\; w_4=\sqrt{r_4}\end{equation}
and then choose  $r_2$ as a new independent variable $y$ which gives rise to the following three equations:
\begin{equation} \left\{
\begin{array}{rcl}
    \displaystyle\der{r_1}{y}&=&-\displaystyle\frac{1}{y}\sqrt{\frac{r_3 (1-r_1^2)}{r_{4}}},\\ [0.37cm]
    \displaystyle\der{r_3}{y} &=& -\displaystyle\frac{2\sqrt{r_{3}}N_{3}}{ y\sqrt{1-r1^2}\sqrt{r_4}r_1 D},\\ [0.37cm]
    \displaystyle\der{r_4}{y} &=& \displaystyle\frac{8N_{4}}{\left( y^{4}-1 \right)D},
\end{array}  \right. \label{HDIVCeqred}
\end{equation}
where:
\begin{align}
    N_{3} &
    \begin{aligned}[t]
            &=
    16 {y}^{4} \left( c_{{1}}-c_{{3}} \right)  \left( a+2 \right) {r_{{1}}}^{4}
    + \left[  \left( 14 c_{{1}}+8 c_{{2}}+10 c_{{3}} \right) a+36 c_{{1}}-16 c_{{2}}-20 c_{{3}} \right] {y}^{4}
    \\
    &+ \big[ 2  \left( y-1 \right) ^{2} \left( y+1
 \right) ^{2} \left( {y}^{2}+1 \right) ^{2} \left( a-2 \right) r_{{3}}
+2  \left( y-1 \right) ^{2} \left( y+1 \right) ^{2} \left( {y}^{2}+1
 \right) ^{2} \left( a-2 \right) r_{{4}}\big.
        \\
        &\left.+8 {y}^{2} \left(  \left( c_{
{2}}-c_{{3}} \right)  \left( a-2 \right) {y}^{4}+ \left(  \left( -4 c
_{{1}}-2 c_{{2}}-2 c_{{3}} \right) a-8 c_{{1}}+4 c_{{2}}+4 c_{{3}
} \right) {y}^{2}+ \left( c_{{2}}-c_{{3}} \right)  \left( a-2 \right)
 \right)  \right] {r_{{1}}}^{2}
        \\
        &- \left( y-1 \right) ^{2} \left( y+1 \right) ^{2} \left( {y}^{2}+1 \right) ^{2} \left( a-2 \right) r_{{3}}
- \left( y-1 \right) ^{2} \left( y+1 \right) ^{2} \left( {y}^{2}+1
 \right) ^{2} \left( a-2 \right) r_{{4}}
 \\
 &+ \left( c_{{1}}-c_{{3}}
 \right)  \left( a-2 \right) {y}^{8}-4  \left( c_{{2}}-c_{{3}}
 \right)  \left( a-2 \right) {y}^{6}
 -4  \left( c_{{2}}-c_{{3}} \right)  \left( a-2
 \right) {y}^{2}+ \left( c_{{1}}-c_{{3}} \right)  \left( a-2 \right),
    \end{aligned}
    \\
    \\
    N_{4} &
    \begin{aligned}[t]
        &=8 {y}^{2} \left[  \left( {y}^{4}+1 \right) r_{{3}}+
 \left( {y}^{4}+1 \right) r_{{4}}+2 {y}^{2} \left( c_{{2}}-c_{{3}}
 \right)  \right]  \left( a+2 \right) {r_{{1}}}^{4}
        \\
        &+8 {y}^{2} \left[
            -\left( {y}^{4}+1 \right) \left(r_{{3}}+ r_{{4}}\right)+
 \left( c_{{1}}-c_{{3}} \right) {y}^{4}+ \left( -2 c_{{2}}+2 c_{{3}}
 \right) {y}^{2}+c_{{1}}-c_{{3}} \right]  \left( a+2 \right) {r_{{1}}}^{2}
        \\
        &+ \left( c_{{2}}-c_{{3}} \right)  \left( a-2 \right) {y}^{8}+
 \left[  \left( -8 c_{{1}}-4 c_{{2}}-4 c_{{3}} \right) a-16 c_{{1}
 }+8 c_{{2}}+8 c_{{3}} \right] {y}^{6}+6  \left( c_{{2}}-c_{{3}}
 \right)  \left( a-2 \right) {y}^{4}
        \\
        &+ \left[  \left( -8 c_{{1}}-4 c_
        {{2}}-4 c_{{3}} \right) a-16 c_{{1}}+8 c_{{2}}+8 c_{{3}} \right] {
y}^{2}+ \left( c_{{2}}-c_{{3}} \right)  \left( a-2 \right),
    \end{aligned}
    \\
    D &= -16 {y}^{4} \left( a+2 \right) {r_{{1}}}
^{4}+16 {y}^{4} \left( a+2 \right) {r_{{1}}}^{2}+ \left( {y}^{4}-1 \right) ^{2} \left( a-2
\right).
    \label{eq:NDHIVC}
\end{align}
From the Hamiltonian ${\cal H}_{DIVC}$, i.e.
\begin{equation}
    {\cal H}_{DIVC}=
    \frac {\left\{
        \begin{gathered}
            \left[  4\left( {y}^{4}-1 \right)^{2} \left(r_{{3}}+ r_{{4}}\right)
            +16 {y}^{2} \left( \left( c_{{2}}-c_{{3}} \right) \left({y}^{4}+1\right)
        - 2\left(c_{{2}}+c_{{3}}\right) {y}^{2}\right)  \right] r_{{1}}^{4}
                -4 c_{1} \left({y}^{4}-1\right)^{2}
                \\
                -\left[  4\left( {y}^{4}-1 \right)^{2} \left(r_{{3}}+r_{{4}}\right)
                 - 4\left( c_{{1}}-c_{{3}}\right) \left({y}^{8}+1\right)
                 +16 \left(c_{{2}}- c_{{3}} \right) {y}^{2}\left( y^{4}+1 \right)
             +8\left(c_{{1}}-4 c_{{2}}-5 c_{{3}} \right) {y}^{4} \right] r_{{1}}^{2}
        \end{gathered}\right\}
}{\left( y^{4}-1 \right) ^{2} \left( a-2 \right)
    -16 {y}^{4} \left( a+2 \right)r_{{1}}^{2}\left( {r_{{1}}}^{2}-1\right)
}
    =h_0,
\end{equation}
we can derive:
\begin{equation}
r_3=-r_4-\displaystyle\frac{c_1}{r_1^2}-\frac{4y^2c_2}{(y^2+1)^2}+\frac{(4r_1^2y^2+y^4-6y^2+1)c_3}{(y^2-1)^2(r_1^2-1)}+\frac{(a-2)h_0}{4r_1^2(r_1^2-1)}-\frac{4(a+2)h_0y^4}{(y^4-1)^2},
\end{equation}
with $h_0$ an arbitrary constant. Consequently, the third equation in \eqref{HDIVCeqred} becomes:
\begin{equation}
    \displaystyle\der{r_4}{y}=8y\frac{(y-1)^4 (y+1)^4 c_2-(y^2+1)^4 c_3+2 h_0 y^2 (y^4+1) (a+2)}{\left( y^{4}-1 \right)^{3}},
\end{equation}
 that can be easily integrated, i.e.:
 \begin{equation}
     r_4=4c_2\frac{(y^2+1)^2-y^2}{(y^2+1)^2} +4c_3\frac{y^4-y^2+1}{(y^{2}-1)^2}-2(a+2)h_0\frac{y^8+1}{(y^{4}-1)^2}+w_{0}
\end{equation}
 with $w_0$ an arbitrary constant. If we introduce new constants $C_2,C_3,C_1,A$ as follows :
 \begin{equation}
 c_2=C_2+c_3,\;  c_3= \frac{4h_0 - w_0 - C_3 - 4C_2 + 2 a h_0}{8},\;c_1=\frac{ - 2a h_0 - C_1 + 4 h_0}{8},\; a= \frac{4C_2 + 9C_3 + w_0- A- C_1}{4 h_0},
 \end{equation}
  then we are left with the following simplified expression of the first equation in \eqref{HDIVCeqred}:
 \begin{equation}
 \displaystyle\der{r_1}{y}=\displaystyle\frac{y^4-1}{2yr_1}\sqrt{\displaystyle\frac{8C_3r_1^4-Ar_1^2-C_1}{2C_3y^8+8C_2y^6+4w_0y^4+8C_2y^2+2C_3}},
  \end{equation}
 which could be solved by quadratures. However, if we solve it with respect to $A$, and derive once by $y$, then a second-order equation is obtained, that admits a three-dimensional Lie symmetry algebra $\Sl(2,\R)$, and as a particular case  if $C_1$ is equal to zero then it is linearizable since it admits an eight-dimensional Lie symmetry algebra $\Sl(3,\R)$. If we solve this second-order equation with respect to $C_1$, and derive once by $y$,  then the following third-order equation is obtained ($r_1\equiv u$):
   \begin{equation}
       \begin{aligned}
           \displaystyle\der[3]{u}{y} &=
           -\frac{3}{u} \der{u}{y}\der[2]{u}{y}
           -\frac{3 \left[C_3 y^{12}-\left(5 C_3+2 w_0\right) y^8-24 C_2 y^6-\left(3 C_3+6 w_0\right) y^4-8 C_2 y^2-C_3\right]}{%
           y (y^{4}-1) (C_3 y^8+4 C_2 y^6+2 w_0 y^4+4 C_2 y^2+C_3)}
           \left[
               \der[2]{u}{y}+\frac{1}{u}\left( \der[2]{u}{y} \right)^{2}
           \right]
                      \\
           &+\frac{3 \left[C_3 y^{16}-\left(6 C_3+2 w_0\right) y^{12}-40 C_2 y^{10}-2\left(7 C_3+10 w_0\right) y^8-80 C_2 y^6-2\left(7 C_3+5 w_0\right) y^4-8 C_2 y^2+C_3\right]}{%
           y^2 (y^{4}-1)^2 (C_3 y^8+4 C_2 y^6+2 w_0 y^4+4 C_2 y^2+C_3)}
           \der{u}{y},
       \end{aligned}
   \label{HIVCeqo3}
   \end{equation}
which is linearizable since it admits a seven-dimensional Lie symmetry algebra. In fact, the two-dimensional abelian intransitive subalgebra generated by the two operators
\begin{equation}
 u^{-1}\partial_{u},\quad \quad \displaystyle\frac{C_3+2C_2y^2+C_3y^4}{uy^2}\partial_u
\end{equation}
when put into the canonical form $\partial_{\tilde u},\tilde y\partial_{\tilde u}$ yield the new dependent and independent variables, i.e. \begin{equation}\tilde u=\frac{u^2}{2},\quad\quad \tilde y= \frac{C_3+2C_2y^2+C_3y^4}{y^2},\end{equation} that transform equation \eqref{HIVCeqo3} into the linear equation
\begin{equation}\displaystyle\der[3]{\tilde u}{\tilde y}=\displaystyle\frac{3\tilde y}{\omega^2-\tilde y^2}\displaystyle\der[2]{\tilde u}{\tilde y},
\end{equation}
with $\omega^2=2(2C_2^2+C_3^2-C_3w_0)$.

\subsection{Case (D)}
 The
Hamiltonian
\begin{equation}
{\cal H}_{DIVD}=-\displaystyle 4w_1^2w_2^2\frac{w_3^2+w_4^2+d\left(\frac{1}{w_1^2}+\frac{1}{w_2^2}\right)}{(a+2)w_1^2+(a-2)w_2^2}
\end{equation}
is a subcase of Hamiltonian ${\cal H}_{DIVA}$, with $a_1=a_3=0$ and $a_2=d$. Consequently, its corresponding Hamiltonian equations, i.e.:
\begin{equation} \left\{
\begin{array}{rcl}
\dot w_1&=&-\displaystyle\frac{8w_1^2w_2^2w_3}{a(w_1^2 + w_2^2) + 2(w_1^2-w_2^2)},\\ [0.37cm]
\dot w_2&=& -\displaystyle\frac{8w_1^2w_2^2w_4}{a(w_1^2 + w_2^2) + 2(w_1^2-w_2^2)},\\ [0.37cm]
\dot w_3 &=& \displaystyle{\frac{8w_1w_2^2}{\left(a(w_1^2 + w_2^2) + 2(w_1^2-w_2^2)\right)^2}}\Big(aw_2^2(w_3^2 +w_4^2)  - 4d - 2w_2^2(w_3^2+w_4^2)\Big),\\[0.37cm]
\dot w_4 &=& \displaystyle{\frac{8w_1^2w_2}{\left(a(w_1^2 + w_2^2) + 2(w_1^2-w_2^2)\right)^2}}\Big(aw_1^2(w_3^2 +w_4^2)+ 4d + 2w_1^2(w_3^2+w_4^2)\Big).
\end{array}  \right. \label{HDIVDeq}
\end{equation}
can be reduced to the following system of three equations, after making the substitutions \eqref{allroots} and choosing  $r_2$ as a new independent variable $y$:
\begin{equation} \left\{
\begin{array}{rcl}
 \displaystyle\der{r_1}{y}&=&\sqrt{\displaystyle\frac{r_1r_3}{yr_4}},\\ [0.35cm]
\displaystyle\der{r_3}{y} &=&-\displaystyle{\frac{\sqrt{r_1r_3}}{r_1\sqrt{yr_4}\left(a(r_1+y) + 2(r_1-y)\right)}}\Big(ay(r_3 +r_4) - 4d  - 2y(r_3+r_4)\Big),\\[0.37cm]
\displaystyle\der{r_4}{y} &=&\displaystyle{\frac{1}{y\left(a(r_1+y) + 2(r_1-y)\right)}}\Big(ar_1(r_3 +r_4)+ 4d+ 2r_1(r_3+r_4)\Big).
\end{array}  \right. \label{HDIVDeqred}
\end{equation}
Then,
\begin{equation}
r_3=-\displaystyle{\frac{4r_1r_4y + ((a+ 2)r_1 + (a - 2)y)h_0 + 4d(r_1 + y)}{4r_1y}},
\end{equation}
and
 \begin{equation}
r_4=\displaystyle\frac{4r_0y-(a+2)h_0-4d}{4y},
\end{equation}
and the first equation in \eqref{HDIVDeqred} becomes:
\begin{equation}
 \displaystyle\der{r_1}{y}=\sqrt{\displaystyle\frac{(2-a)h_0 - 4d  - 4r_1r_0)}{-(a+2)h_0 - 4 d + 4 r_0 y}},\label{HIVD1ode}
 \end{equation}
 which could be easily solved by quadratures. However, if we make the simplifying substitution $d=D-(a+2)h_0/4$, solve the first-order equation \eqref{HIVD1ode} with respect to $h_0$, and derive once with respect to $y$, then the following linear second-order equation is obtained:
 \begin{equation}
 2(r_0y- D)\displaystyle\der[2]{r_1}{y} + r_0 \displaystyle\der{r_1}{y} + r_0=0.
 \end{equation}

\section{Conclusions}
In this paper, nineteen classical superintegrable system in two-dimensional non-Euclidean spaces are shown to possess hidden symmetries leading to linearity. This fulfills the conjecture that we made in \cite{GN_supint}, namely that all classical superintegrable system in two-dimensional space hide linearity regardless of the separation of variables of the corresponding Hamilton-Jacobi equation, and of the order of the first integrals.

In some cases, we have used the Hamiltonian in order to derive one of the two momenta as function of the other momentum and coordinates. None of the other two known first integrals have been used. In other cases, one of the equation of the Hamiltonian system could be integrated by quadrature, and that was all we needed in order to then find the hidden symmetries leading to linear equation of either second or third order.

As we stated in \cite{GN_supint}, it remains an open-problem to see if linear equations are hidden in (maximally?) superintegrable systems in $N>2$ dimensions, regardless of the separability of the corresponding Hamilon-Jacobi equation, and the degree of the known first integrals.

\section*{Acknowledgments}
GG acknowledges support by Prof. N. Joshi and A/Prof. M. Radnovi\v{c}'s grant DP200100210 from the Australian Research Council.

MCN acknowledges the partial support of University of Perugia through {\em Fondi Ricerca di Base 2018}.

\end{document}